# Quantum Dot Colloidosomes as Triggerable Microlasers

Cristian Gonzalez[1], Saranya Subramanian,[4] Marco Reale,[4,6] Giuseppe Soligno,[5] Ilia Geints[3], Siyuan Yin[1], Claire Y. Kang[1], Ricky Ronquillo,[1] Gary Chen[1], Marco Cannas,[4,6] Cherie R. Kagan[1,2,3], Alice Sciortino[4], Michael Engel,[5] Fabrizio Messina[4], Christopher B. Murray[1,2]*, Emanuele Marino[4]*

[1] Department of Chemistry, [2] Department of Materials Science & Engineering, and [3] Department of Electrical & Systems Engineering, University of Pennsylvania, Philadelphia (PA), USA.

[4] Department of Physics and Chemistry – Emilio Segrè, University of Palermo, Palermo, Italy

[5]Institute for Multiscale Simulation, IZNF, Friedrich-Alexander-Universität Erlangen-Nürnberg, 91058 Erlangen, Germany.

[6]INSTM – National Interuniversity Consortium of Materials Science and Technology, Florence, Italy

*Corresponding author: emanuele.marino@unipa.it, cbmurray@sas.upenn.edu.

ORCID:

Cristian Gonzalez: 0000-0001-7639-6193

Saranya Subramanian: 0009-0002-9618-3288

Marco Reale: 0000-0001-6849-9061

Giuseppe Soligno: 0000-0003-2360-2082

Ilia Geints: 0000-0002-5061-5885

Siyuan Yin: 0009-0002-3766-8568

Claire Y. Kang: 0009-0006-3982-1748

Ricardo Ronquillo: 0009-0004-8855-1647

Gary Chen: 0009-0001-3503-1654

Marco Cannas: 0000-0001-8236-5043

Cherie R. Kagan: 0000-0001-6540-2009

Alice Sciortino: 0000-0001-8361-3002

Michael Engel: 0000-0002-7031-3825

Fabrizio Messina: 0000-0002-2130-0120

Christopher B. Murray: 0000-0002-3491-122X

Emanuele Marino: 0000-0002-0793-9796

## Abstract

Quantum dot (QD) supraparticle lasers are promising platforms for microscale light sources and photonic labeling, yet their optical properties are set during assembly. Here we introduce QD colloidosomes, liquid-core/solid-shell supraparticles that combine whispering-gallery-mode optical feedback with stimulus-triggered structural collapse. Intact QD colloidosomes show cavity-defined lasing with fluence thresholds of ~2.8 mJ/cm$^2$ and linewidths of 2.2–3.7 nm, demonstrating efficient light trapping without a solid core. We identify shell continuity as a critical determinant for whispering-gallery feedback and lasing by exploring a morphological continuum of suprastructures comprising solid supraparticles, colloidosomes, and microporous hollow shells. Intact colloidosomes support narrow cavity modes, whereas porous shells remain broadband even at high pump fluence. We show that colloidosomes can be driven to rupture and release payload through well-defined pathways, including meniscus-driven capillary failure, uniform heating, and localized near-infrared activation, thereby reconfiguring their optical response from cavity-defined lasing to broadband emission. These results establish QD colloidosomes as reconfigurable microlasers that couple optical-state switching and triggered release within a single self-assembled platform.

## Introduction

The miniaturization and control of light sources are central challenges at the intersection of nanophotonics, materials science, and bioimaging. Colloidal semiconductor nanocrystals, known as quantum dots (QDs), have emerged as ideal building blocks for compact photonic systems, owing to their efficient, size-tunable photoluminescence, exceptional photostability, and compatibility with low-temperature colloidal processing. Accordingly, self-assembly has become a powerful tool to fabricate integrated optical devices. For instance, assembling QDs into densely packed spherical suprastructures, known as supraparticles, enables the emergence of whispering-gallery modes (WGMs).[1, 2] These photonic modes provide optical feedback, resulting in light amplification and lasing under optical excitation. In these self-assembled microlasers, QDs provide optical gain while the supraparticle geometry defines the optical cavity, resulting in high quality factors ($Q > 1000$) and low lasing thresholds without an external resonator.[3] As a result, microscale lasers are increasingly attractive as tunable light sources and as photonic labels for sensing, imaging, and tagging.[4]

Despite these advances, self-assembled QD microlasers generally lack built-in structural stimulus responsiveness after assembly. The optical response of a microcavity is largely determined by its size, composition, and geometry, offering only modest post-fabrication tunability through indirect mechanisms such as temperature-dependent cavity shifts, carrier-induced refractive index changes, or state-filling-driven redistribution of optical gain.[3, 5, 6] Nanophotonic systems with stimulus-responsive cavity architectures could couple irreversible optical-state switching to triggered payload release, extending the functionality of conventional supraparticle lasers.

Stimuli-responsive microcontainers provide a complementary route to dynamic microscale function: lipid vesicles and polymer capsules can release encapsulated contents or change permeability in response to heat,[7] light,[8] or chemical cues,[9] and plasmonic nanoparticle-based photothermal actuators enable laser-triggered rupture of individual microcapsules.[10] More broadly, colloidosomes, capsule-like structures formed by colloidal particles assembled and jammed at liquid-liquid interfaces, represent a class of particle-stabilized microcontainers.[11] However, these systems lack the photonic function, optical gain, and cavity feedback typical of semiconductor microlasers. Conversely, QD supraparticle lasers can exhibit impressive optical performance but lack an intrinsic mechanism for stimulus-triggered reconfiguration or release. Bridging this gap between high-performance optical devices and stimulus-responsive materials remains an open challenge.

Here, we introduce QD colloidosome supraparticles that couple WGM-assisted lasing with stimulus-triggered shell rupture and release of an encapsulated liquid core. Each microparticle consists of a thin shell of assembled QDs enclosing a microscopic alkane-rich liquid core containing dispersed QDs. In this state, the colloidosome supports optical feedback through WGMs, leading to amplified spontaneous emission and lasing. However, meniscus-driven capillary failure, uniform heating, and focused near-infrared (NIR) irradiation can induce controlled shell rupture and colloidosome collapse. This structural transformation reconfigures the optical response from cavity-defined lasing to broadband emission while enabling rapid release of the QD core material. Across solid supraparticles, closed colloidosomes, and microporous hollow shells, we identify shell continuity as the key structural variable linking optical feedback, lasing, and triggered release. These results define a strategy for self-assembled semiconductor microcavities where shell integrity acts as a common control parameter for light generation, optical shutdown, and triggered release.

## Results and Discussion

**Synthesis of quantum dot colloidosomes.** Figure 1A-C illustrates the synthesis of quantum dot (QD) colloidosomes. We adapt microfluidic "source-sink" emulsion templated assembly from our previous work on solid supraparticles.[12] We use a glass microfluidic device where an oil stream (0.01–0.05 vol % of C12-C18 n-alkane in toluene) is sheared by a continuous aqueous phase (20 mM SDS in water) to generate monodisperse oil-in-water "source" droplets (Figure 1A). These source droplets contain oleate-capped core/shell CdSe/CdS QDs (9 nm diameter, 2 mg $mL^{-1}$). After generation, the source droplets are mixed with smaller "sink" droplets (~60 nm) consisting of pure hexadecane (Figure 1B). The source-sink system evolves through the preferential transfer of toluene from source droplets to sink droplets, enabled by the large asymmetry ($10^7$–$10^8$)[13] in aqueous solubility between toluene and hexadecane.

This preferential toluene transfer drives the assembly of QDs into suprastructures that are denser than the continuous phase, enabling isolation by mild centrifugation (~100 g) in glass vials and redispersion in 20 mM SDS in water. Figure 1D shows a representative scanning electron microscopy (SEM) image of fully dried suprastructures obtained when using an initial volume fraction of 0.03% hexadecane. After drying, the suprastructures are approximately spherical, 9.58 ± 2.48 μm in diameter, and exhibit a hollow morphology enclosed by a thin, solid shell with an average thickness of 275 ± 50 nm. This morphology deviates significantly from that of solid supraparticles studied by our group and others,[1, 2, 14-18] consisting of densely packed QD assemblies. The morphology observed here indicates that, prior to drying, the suprastructures contained an alkane-rich liquid core enclosed by a solid shell of assembled QDs, as shown in Figure 1E. In the soft matter community, this type of architecture is known as *colloidosome*.[11, 19, 20] The liquid core enclosed within a colloidosome introduces the potential for cargo release upon shell disruption. In our specific system, the combination of a QD shell and liquid core raises the possibility of coupling optical function to stimulus-triggered rupture and release, as shown schematically in Figure 1F.

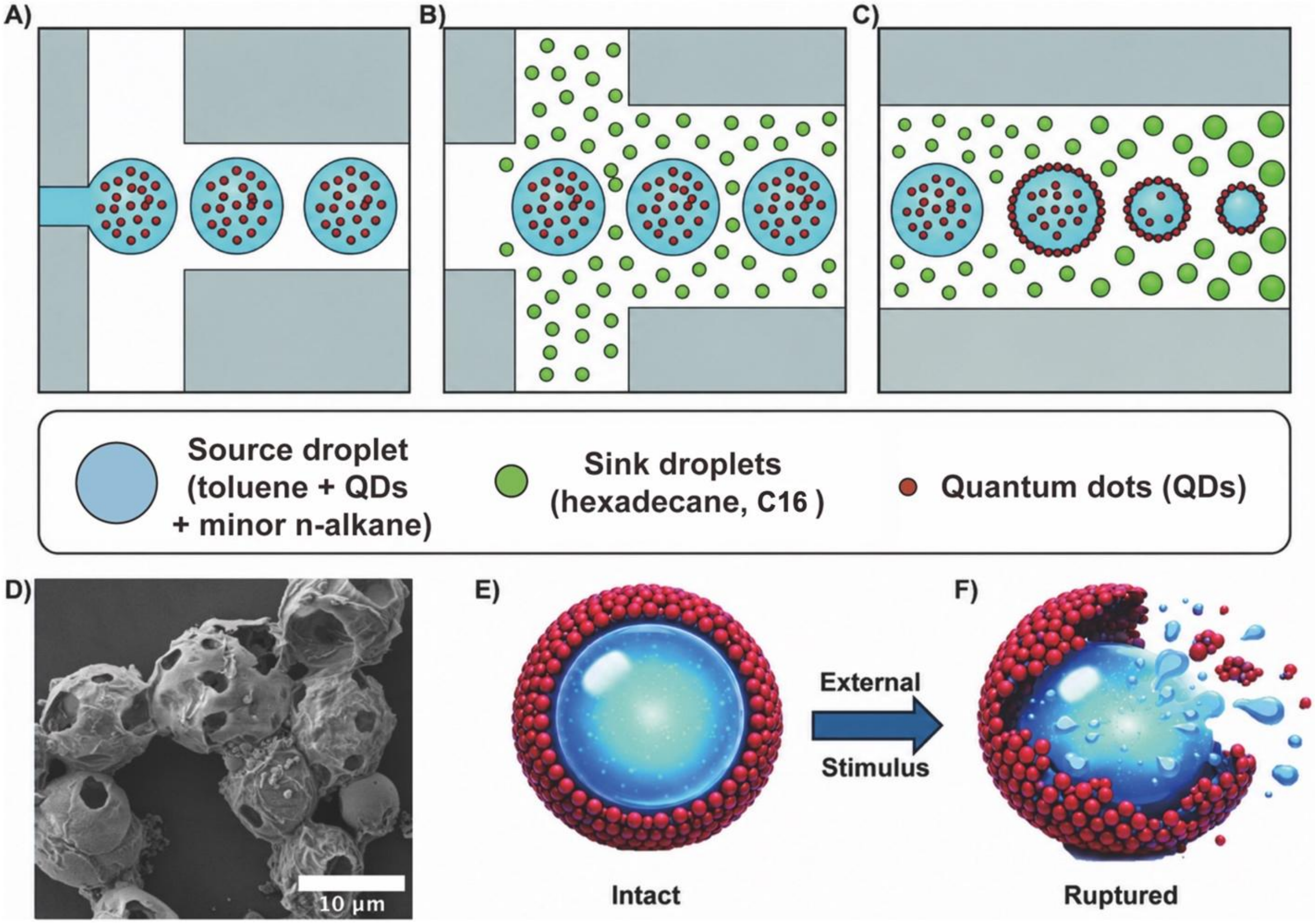

**Figure 1. Microfluidic source–sink emulsion-templated assembly of quantum dot (QD) colloidosomes. (A)** Schematic generation of monodisperse oil-in-water microfluidic "source" droplets containing oleate-capped CdSe/CdS QDs dispersed in toluene with a minor fraction of n-alkane. **(B)** The source droplets are mixed with an excess "sink" emulsion of smaller droplets of pure hexadecane. **(C)** Toluene transfer from the source to the sink droplets increases QD concentration and drives assembly at the droplet interface. **(D)** Representative scanning electron micrograph of QD colloidosomes following drying-induced shell rupture. **(E)** Conceptual rendering of an intact liquid-core/solid-shell QD colloidosome. **(F)** Conceptual rendering of colloidosome rupture by external stimuli.

**Interfacial model for colloidosome formation.** Understanding the mechanism of colloidosome formation is key to enabling rational control over structure-property relationships. Figure 2A defines the interfacial geometry for a QD at an oil–water interface. The change in interfacial energy upon transferring a QD from the oil phase to the oil–water interface can be written as:

$$E = \gamma_{ow}(S - S_0) + \gamma_{no}(W_o - \Sigma) + \gamma_{nw}W_w \quad (1),$$

where $\gamma_{ow}$, $\gamma_{no}$, and $\gamma_{nw}$ are the interfacial tensions between oil-water, QD-oil, and QD-water, respectively; $S$ is the oil-water interfacial area in the presence of the QDs; $S_0$ is the oil-water interfacial area in the absence of the QDs; $W_o$ and $W_w$ are the QD surface areas wetted by oil and water, respectively; and $\Sigma \equiv W_o + W_w$ is the total surface area of the QD. The reference level is defined such that $E = 0$ corresponds to the desorbed QD, fully immersed in the oil phase.

These interfacial tensions determine effective wettability via Young's law,

$$\cos\theta \equiv \frac{\gamma_{nw} - \gamma_{no}}{\gamma_{ow}} \quad (2),$$

where $\theta$ is the contact angle measured inside the oil phase. For $\cos\theta > 1$, the system lies in the oil-wetting regime, and the QD remains immersed in the oil phase. For $\cos\theta < -1$, the QD would instead be fully wetted by the aqueous phase. When $-1 \leq \cos\theta \leq 1$, the system enters the partial-wetting regime, in which interfacial adsorption becomes favorable, and the QD adsorbs to the interface.

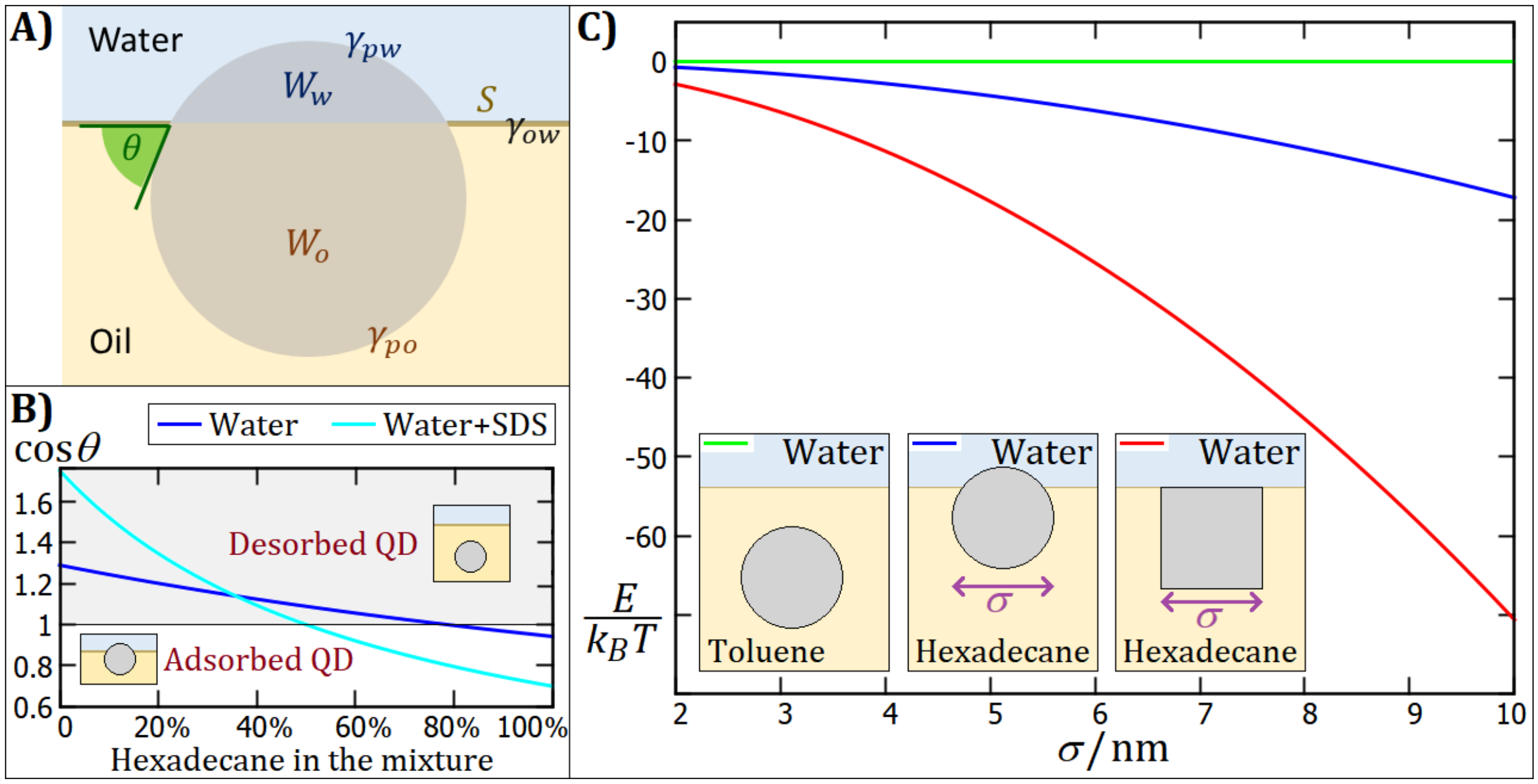

**Figure 2. Thermodynamics of interfacial adsorption of oleate-capped quantum dots (QDs) during solvent exchange**. (A) Schematic defining interfacial areas and surface tensions for a QD adsorbed at an oil–water interface; $\theta$ is Young's contact angle measured through the oil phase. (B) Predicted values of $\cos\theta$ for an oleate-capped QD at the interface between water and a mixture of toluene and hexadecane versus hexadecane volume fraction, shown for pure water and for aqueous solutions of sodium dodecyl sulfate (SDS) exceeding the critical micelle concentration. (C) Predicted adsorption energy $E$ in units of $k_BT$ at the hexadecane–water interface under representative SDS conditions ($cos\,\theta = 0.7, \gamma_{ow} = 10$mN m$^{-1}$), comparing spherical (diameter $\sigma$) and cubic (side $\sigma$) particle geometries; for the toluene–water interface, the equilibrium state corresponds to desorption ($E = 0$).

We use reported values of interfacial tension: for toluene, $\gamma_{ow} \approx 38\ \mathrm{mN\ m^{-1}}$ with pure water and $\gamma_{ow} \approx 4\ \mathrm{mN\ m^{-1}}$ with aqueous SDS solutions near or above the critical micelle concentration (CMC);[21, 22] for hexadecane, $\gamma_{ow} \approx 52\ \mathrm{mN\ m^{-1}}$ with pure water and $\gamma_{ow} \approx 10\ \mathrm{mN\ m^{-1}}$ with aqueous SDS solutions near/above the CMC.[23] Since the QDs are capped with oleates soluble in both toluene and hexadecane, we take $\gamma_{no} \approx 0$ to be small compared to $\gamma_{ow}$ and $\gamma_{nw}$. We estimate the QD-water interfacial tension, $\gamma_{nw}$ by approximating the interfacial tension between the alkane-like tail of the oleate and water using hexane–water as a proxy, resulting in $\gamma_{nw} \approx 49\ \mathrm{mN\ m^{-1}}$ for pure water and $\gamma_{nw} \approx 7\ \mathrm{mN\ m^{-1}}$ for aqueous SDS solutions near/above the CMC. These estimates place the QDs near the oil-wetting/partial-wetting boundary, so modest changes in $\gamma_{ow}$ or $\gamma_{nw}$ can shift the system between dispersed and interfacially adsorbed states.

Using reported values for interfacial tension (see SI), Eq. (2) predicts the full-wetting regime for the toluene–water–QD system ($\cos\theta > 1$): Oleate-capped QDs do not favor adsorption at the toluene-water interface and remain confined to the oil phase ($E = 0$). In contrast, replacing toluene with hexadecane shifts the system into the partial-wetting regime: Eq. (2) yields $\cos\theta \approx 0.94$ in pure water and $\approx 0.70$ in the presence of SDS, consistent with stable adsorption of QDs to the hexadecane-water interface ($E < 0$).

We extend this analysis to the mixed toluene–hexadecane oil phase present in our experiments (Figure 2B). As a first approximation, linear interpolation of the oil–water interfacial tension, $\gamma_{ow}$, between pure toluene and hexadecane predicts a composition-dependent adsorption threshold. QDs are predicted to adsorb to the interface once the system crosses from the oil-wetting regime into the partial-wetting regime only when the hexadecane volume fraction exceeds ~80% in pure water and ~50% in SDS solutions near or above the critical micelle concentration. These results are consistent with our experimental observations: QDs within toluene-rich droplets are well dispersed but become adsorbed at the liquid–liquid interface as the droplet becomes alkane-rich.

Figure 2C shows the equilibrium adsorption energy $E$ as a function of QD size for spherical and cubic geometries; see SI for details.[24] Our QDs adopt morphologies intermediate between these limits, as they consist of a CdSe zincblende core coated with a CdS epitaxial shell (Fig. S1). Under conditions relevant to our experiments, the adsorption energy is strongly negative, with adsorption energies of tens of $k_B T$ in magnitude that increase with QD size. For 9 nm QDs used in this work, $E$ reaches -14 $k_B T$ for spherical geometry and -57 $k_B T$ for cubic geometry. This implies that QD adsorption is effectively irreversible, consistent with the formation of the robust colloidosome shell observed in experiments.[25]

**From supraparticles to colloidosomes to microporous hollow shells.** This theoretical framework helps rationalize solid supraparticles and colloidosomes within a single family of suprastructures, with morphologies spanning a continuum accessible through systematic variation of the initial alkane loading, which changes the solvent-exchange pathway and QD formation at the liquid–liquid interface.[26] We probe this structural continuum in experiments by varying the initial loading of long-chain alkane, which directs the assembly pathway from solid supraparticles (no alkane) to liquid-core/solid-shell colloidosomes (intermediate alkane content) and ultimately to microporous hollow shells (high alkane content). We distinguish structures observed in the wet state from those introduced during drying by pairing optical micrographs of particles suspended in aqueous SDS with SEM micrographs acquired after deposition and drying. The accompanying 3D renderings are conceptual guides to the dominant morphology assignments and are not direct volumetric reconstructions of the particle interiors.

At 0% v/v initial alkane content (Figure 3A-C), the optical micrograph shows round, dark, and featureless colloids in water, while SEM reveals largely spherical supraparticles with only small surface pits and no macroscopic pores.[2] This corresponds to the solid-supraparticle limit, in which the suprastructures appear closed before drying and largely retain their spherical morphology afterward.

At 0.02% v/v initial alkane content, the wet-state optical micrograph shows closed particles, whereas dried-state SEM reveals pronounced surface wrinkling, buckling, and partial collapse (Figure 3D–F). Relative to the more shape-retaining no-alkane supraparticle control, this deformation is consistent with a mechanically compliant shell surrounding an internal volume. However, these panels do not directly resolve the composition of an intact particle interior, and Figure 3F is a conceptual morphology assignment rather than a direct three-dimensional reconstruction.

We assign the 0.02% v/v population as liquid-core/solid-shell colloidosomes based on combined evidence rather than on Figures 3D-F alone. The source-sink assembly pathway and interfacial-energy analysis support QD adsorption and assembly at the retained alkane-water interface. Most importantly, particles prepared under the same 0.02% v/v condition undergo abrupt rupture followed by the outward release and spreading of fluorescent QD-containing material under meniscus-driven and thermal triggers (Figures 5-6 and Supporting Movies 2-4). Overall, these observations support the colloidosome assignment for supraparticles prepared at 0.02% v/v initial alkane content.

A striking visual transition emerges at higher initial alkane content. At 0.03% v/v (Figure 3G–I), many suprastructures show visible openings while still immersed in water, and SEM confirms clear pores with thinned rims and partially collapsed shells. This observation indicates that wet-state porosity emerges when the initial oil loading is sufficiently high. At 0.05% v/v (Figure 3J–L), the pores become prevalent: optical images show a large fraction of particles with openings in water, while SEM reveals larger, often connected apertures and frequent drying-associated shell-collapse features.

These results show that the initial alkane loading shifts the architecture of the assembled suprastructures from closed, payload-bearing shells to porous shells. We propose that increasing the concentration of long-chain alkanes promotes attractive interactions between dispersed QDs, ultimately leading to pore formation. In this regime, QDs may preferentially associate onto particles already adsorbed at the interface rather than occupying bare interfacial area, resulting in an increase in shell thickness at the expense of complete surface coverage. Similar porous morphologies have recently been reported using amphiphilic polymers that spontaneously adsorb to the oil-water interface, indicating that porous-shell formation can arise from incomplete interfacial coverage albeit through distinct mechanisms.[27] In our system, alkane loading thus acts as a structural control parameter that determines whether the colloidosome forms a closed, thin-shelled, payload-bearing microcavity or evolves into a porous, thick-shelled suprastructure.

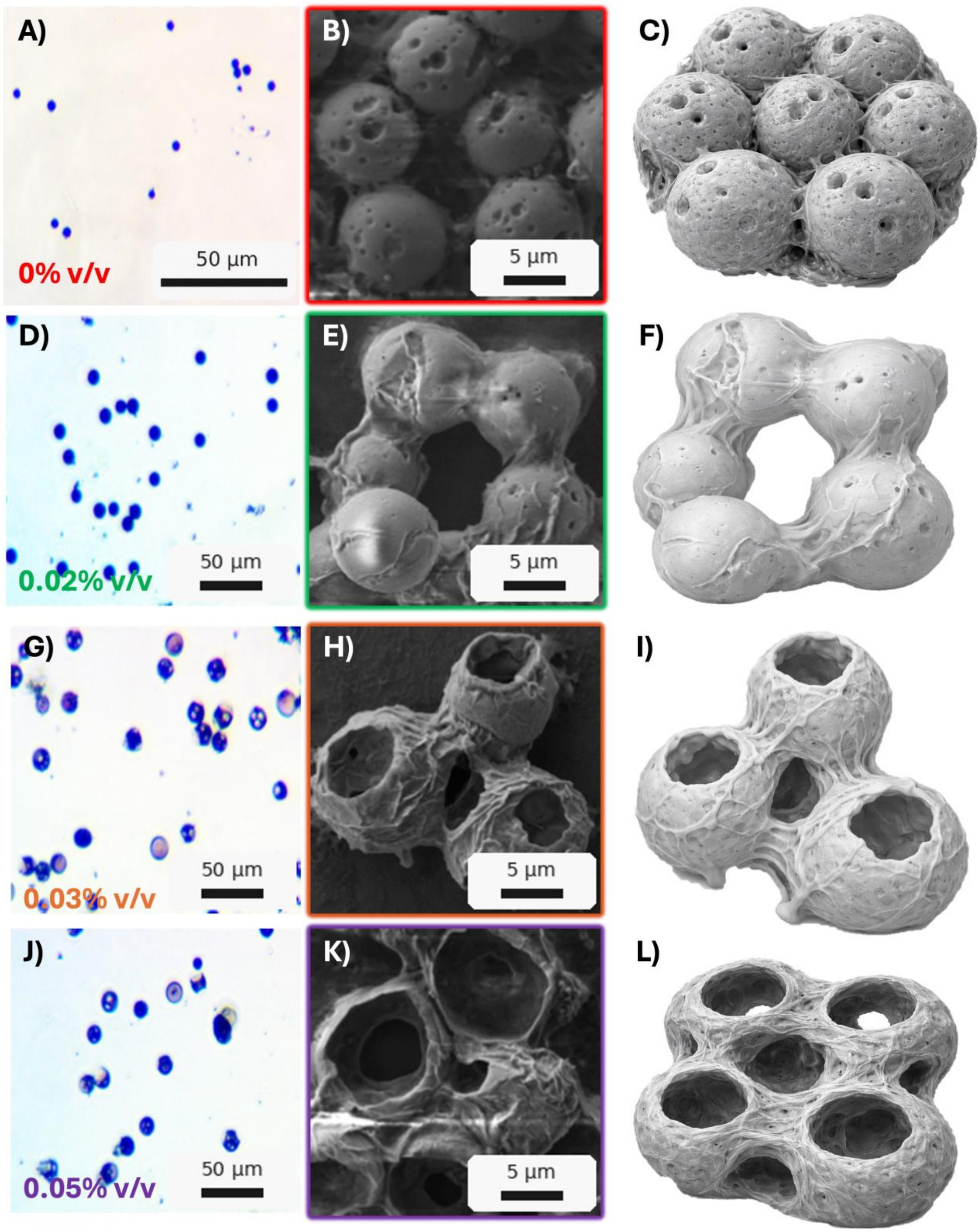

**Figure 3. Experimental mapping of the morphological continuum of quantum dot (QD) suprastructures.** (A,D,G,J) Optical micrographs of suprastructures dispersed in aqueous SDS after preparation (wet state) for increasing initial alkane loading (0, 0.02, 0.03, and 0.05% v/v). (B,E,H,K) Corresponding SEM micrographs after deposition and drying. (C,F,I,L) 3D schematic renders illustrating the dominant morphologies inferred from wet-state optical microscopy and dried-state SEM. At 0-0.02% v/v, particles appear closed in water, while SEM after drying shows only pits/wrinkling. At ≥0.03% v/v, apertures are evident in both wet optical images and dried SEM, indicating that perforations are already present before drying and are more pronounced after drying.

**Quantum Dot Colloidosome Lasers:** The existence of a morphological continuum spanning solid supraparticles, colloidosomes, and microporous shells provides an opportunity to study the relationship between structure and optical function. Microscale dielectric objects with rotational symmetry can confine light along their surface through whispering-gallery modes (WGMs).[28] In solid supraparticles, such confinement provides optical feedback for lasing.[1, 15, 29]
Although hollow-core WGM resonators, ring resonators, and related shell-like microcavities are well established,[30] the extension of this concept to self-assembled QD colloidosome microlasers remains unexplored. In particular, it remains unclear whether a thin, shell-confined QD gain medium surrounding a liquid core can provide sufficient WGM feedback for lasing while simultaneously preserving stimulus-triggered rupture and release functionality. QD colloidosomes therefore provide an ideal platform to test whether shell-confined QD gain and WGM feedback are sufficient to support lasing in a liquid-core, solid-shell colloidal architecture.

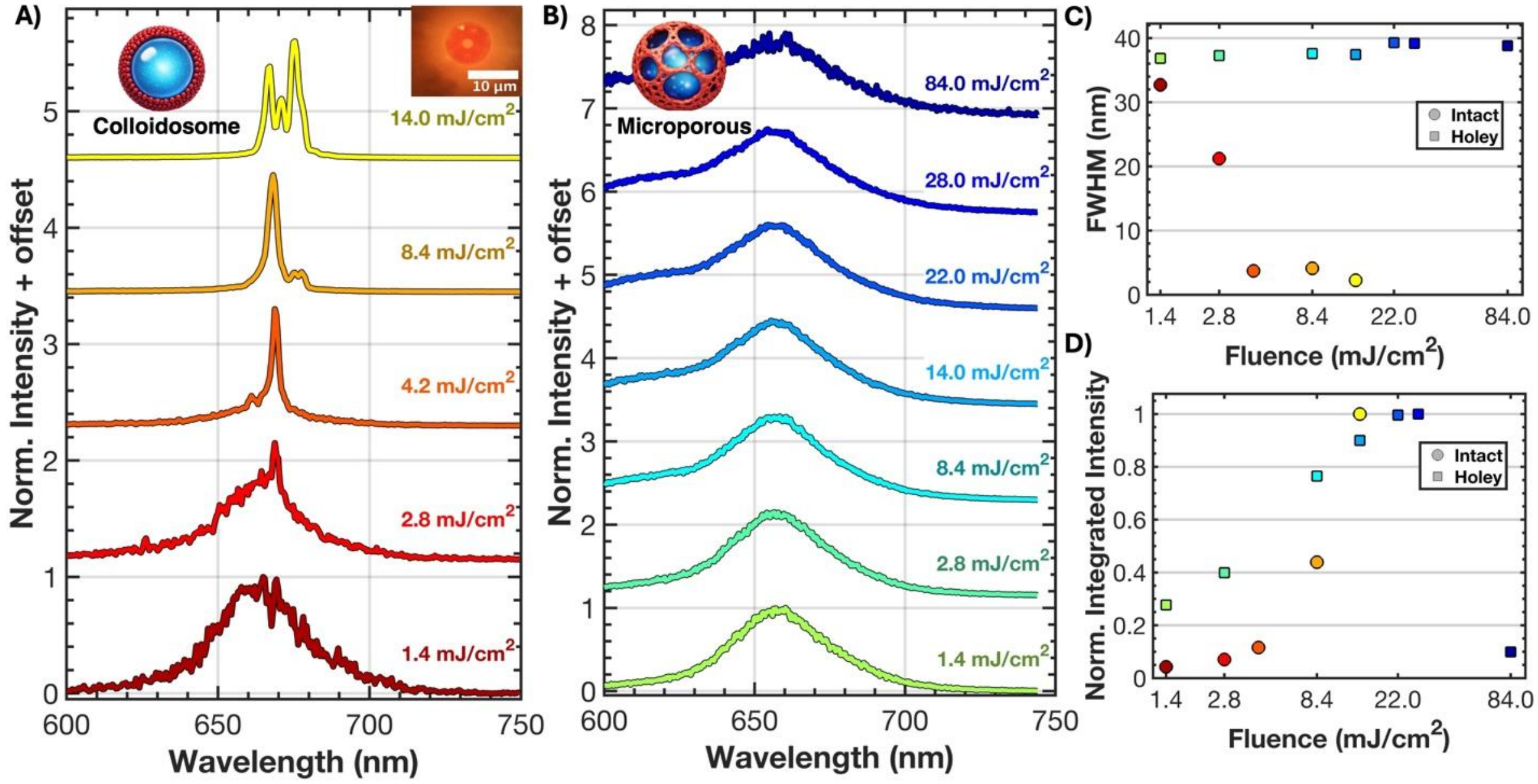

**Figure 4. Fluence-dependent emission and lasing from individual quantum dot (QD) colloidosomes and microporous suprastructures**. (A) Stacked, normalized emission spectra (offset) from an intact colloidosome prepared with 0.02% v/v initial alkane loading**,** showing the evolution from broadband photoluminescence through an ASE-like transition regime to discrete cavity-defined WGM-assisted lasing modes at higher pump fluence. The inset shows an optical micrograph of the colloidosome under pulsed excitation. (B) Stacked, normalized emission spectra (offset) from a microporous suprastructure prepared with 0.03% v/v initial alkane loading, which remains broadband without the emergence of distinct cavity modes across the fluence range. (C) Emission linewidth (FWHM) versus fluence for the colloidosome and microporous suprastructure shown in A and B. (D) Normalized integrated intensity versus fluence for the same two suprastructures. Panels C and D use logarithmic fluence axes. In C-D, marker shape denotes sample type, and marker colors correspond to the spectra and fluences shown in A-B.

We measure emission from a single QD colloidosome dispersed in 20 mM SDS in water using 2 ns laser pulses at 500 nm excitation and 10 Hz repetition rate. The colloidosome shell has a representative thickness of approximately 300 nm, based on SEM measurements of comparable dried structures (275 ± 50 nm; Figure 1D). Figure 4A shows normalized and vertically stacked PL spectra from an intact colloidosome. The PL spectrum evolves from broadband emission at low fluence (1.4 mJ/cm$^2$) to a mixed transition regime at 2.8 mJ/cm$^2$, where ASE-like spectral narrowing begins and narrow features emerge on top of the broad emission background, and finally to cavity-dominated multimode WGM-assisted lasing at higher fluence. At 4.2 mJ/cm$^2$, a single narrow emission line at 668.7 nm dominates the spectrum. At the highest fluence used, 14.0 mJ/cm$^2$, three closely spaced lasing modes emerge near 667.1, 670.9, and 675.3 nm. The lasing modes shift slightly to shorter wavelength with increasing fluence, likely reflecting pump-induced changes in the active cavity, including carrier density dependent refractive index changes and redistribution of net optical gain across the QD emission band; local heating may also perturb the cavity resonance under pulsed excitation.[3, 15, 30]

Figure 4C shows the corresponding spectral narrowing. The colloidosome exhibits a broad PL envelope at low fluence, with an effective full-width at half-maximum (FWHM) of ~28 nm. At 2.8 mJ/cm$^2$, the spectrum contains both broad PL and emerging ASE-like features and is therefore treated as a mixed transition regime. Above this regime, the dominant cavity features narrow sharply, with the 4.2 mJ/cm$^2$ line exhibiting a linewidth of 3.7 nm and the highest-fluence lasing modes exhibiting fitted linewidths of approximately 2.2-3.5 nm (Figure S6).

The fluence-dependent response provides four experimental signatures supporting assignment of the high-fluence optical behavior as WGM-assisted lasing: (i) discrete cavity-defined peaks emerge at higher fluence; (ii) the broad ~28 nm photoluminescence envelope narrows to dominant features with apparent linewidths of ~2.2–3.5 nm; (iii) the log–log integrated-emission curve exhibits a clear slope change near 2.8 mJ/cm$^2$ (Figure S7); and (iv) the microporous control remains broadband and exhibits neither a comparable threshold-like transition nor discrete cavity modes, even at substantially higher pump fluence. Together, these observations distinguish the high-fluence, cavity-dominated regime from the intermediate-fluence, ASE-like transition regime. Because the fluence-dependent measurements were acquired with ~1 nm spectral resolution, fitted linewidths should be interpreted as apparent linewidths of cavity-defined multimode emission and may include contributions from unresolved closely spaced modes, lifted modal degeneracies, and instrumental broadening.

We quantify the sharpness of the above-threshold emission using an active, linewidth-derived quality factor, $Q_{\text{active}} \approx \lambda/\Delta\lambda$, where $\lambda$ is the wavelength of the lasing mode and $\Delta\lambda$ is the FWHM. Using the three dominant lasing peaks at the highest fluence, we obtain $Q_{\text{active}} \approx 239$, 302, and 196, respectively; see Figure S6 for analysis. These values describe the apparent linewidths of the above-threshold cavity-defined emission modes and are not interpreted as intrinsic passive cavity Q factors, because stimulated emission, gain narrowing, multimode overlap, and pump-dependent cavity/gain dynamics can affect the measured linewidth above threshold. To estimate the passive cavity response more directly, we analyzed resonant features in photoluminescence spectra collected under continuous-wave 976 nm near-infrared excitation. Lorentzian fits to the red-side NIR-excited PL resonances give passive quality factors of $Q_{\text{passive}} \approx 140–511$ for modes centered between 689.5 and 724.9 nm (Figure S9). These passive PL-derived Q values are comparable to, and in several cases higher than, the above-threshold active Q values, supporting the presence of resolvable passive cavity resonances while emphasizing that $Q_{\text{active}}$ and $Q_{\text{passive}}$ describe distinct measurement regimes. Interestingly, during excitation at the highest applied pump fluence of 14.0 mJ/cm², the colloidosome abruptly ruptures, leading to the disappearance of discrete cavity modes, consistent with sudden fragmentation of the shell structure and possible destabilization due to photothermal or photocharging effects in the suprastructure (see Supporting Movie 1).

In contrast to the colloidosome studied in Figure 4A, pulsed excitation of microporous suprastructures of the type shown in Figure 3G-I does not produce discrete lasing modes (Figure 4B). Instead, the emission remains broadband over the full fluence range, with no narrow cavity modes emerging up to 84 mJ/cm$^2$. Consistently, Figure 4D shows that the normalized integrated intensity increases gradually with fluence but lacks a clear threshold-like knee, as further highlighted in the log-log integrated-intensity analysis shown in Figure S7. Figure 4C also shows that the emission FWHM remains broad across the entire fluence range, varying only slightly from ~37 to ~39 nm before high-fluence damage. The broader emission of the microporous structure qualitatively resembles the post-rupture PL in Figure 6E. In both cases, disruption of the continuous QD shell suppresses WGM feedback and cavity-mediated spectral selection, leaving emission dominated by the inhomogeneously broadened QD ensemble; heterogeneous QD packing and energy transfer toward lower-energy emitters may further enhance the red side.[36] The narrow emission observed from both intact colloidosomes and fully solid supraparticles further indicates that cavity continuity, rather than whether the core is solid or liquid, controls WGM-mediated spectral selection.[1,15,29]

At the highest applied fluence of 84 mJ/cm², the microporous shell structure becomes unstable and ultimately disintegrates, consistent with photodamage or structural failure. By contrast, the intact colloidosome demonstrates that a liquid core is compatible with WGM-assisted lasing, provided that the QD shell remains continuous and sufficiently spherical. For a representative colloidosome with a 10 µm outer diameter and a 300 nm shell, the shell occupies approximately 17% of the volume of an equivalently sized solid sphere. At comparable QD packing fractions, the colloidosome architecture could therefore reduce QD use by approximately 83% while retaining gain medium near the resonator periphery, where WGM fields are concentrated. The reduced radial gain volume may also reduce the number of supported radial modes relative to a fully solid supraparticle.

To gain insight into the role of colloidosome shell thickness, we performed FEM eigenmode calculations for ideal continuous shells. These calculations show that the same TE-like WGM remains shell-localized over thicknesses of 250–350 nm, with calculated Q values of approximately $5.2 \times 10^2$ (Figure S10). To examine the thinner-shell regime, we performed a complementary FDTD sweep over shell thicknesses from 25 to 350 nm while holding the outer particle radius and material parameters fixed (Figure S11). Representative field profiles show that the calculated fraction of electric-field intensity localized within the QD shell increases from 13.1% at 25 nm to 56.9% at 70 nm and 95.0% at 300 nm (Figure S12). These results reveal a continuous transition from weakly shell-localized, radiatively lossy thin-shell resonances to strongly shell-confined WGMs, demonstrating that a continuous 300 nm shell supports a strongly shell-localized WGM. Because these passive idealized calculations omit wavelength-dependent gain, mode–gain overlap, material absorption, roughness, porosity, and QD-packing disorder, they do not establish a universal optimum or minimum lasing thickness or quantitatively predict the experimental threshold.

Future studies that systematically vary shell thickness and quantify both cavity loss and mode-gain overlap could determine whether an optimal thickness exists that preserves WGM feedback while minimizing the amount of QD material required.

**Colloidosome rupture and payload release by meniscus-driven capillary stress.** The optical properties of QD suprastructures depend critically on their morphology, as shown in the previous section. This coupling between structure and function suggests that controlled disruption of colloidosomes could be used to modulate their optical response on demand. We investigate the effect of a meniscus-driven mechanical trigger by exposing colloidosomes to a receding water–air meniscus, as schematically shown in Figure 5A. As the drying front reaches the dark spheres, fluorescent material is abruptly released, accompanied by the emergence of a bright center and a dark halo. This observation is consistent with the rupture of the colloidosome shell, resulting in the release of QD-containing alkane-rich liquid.

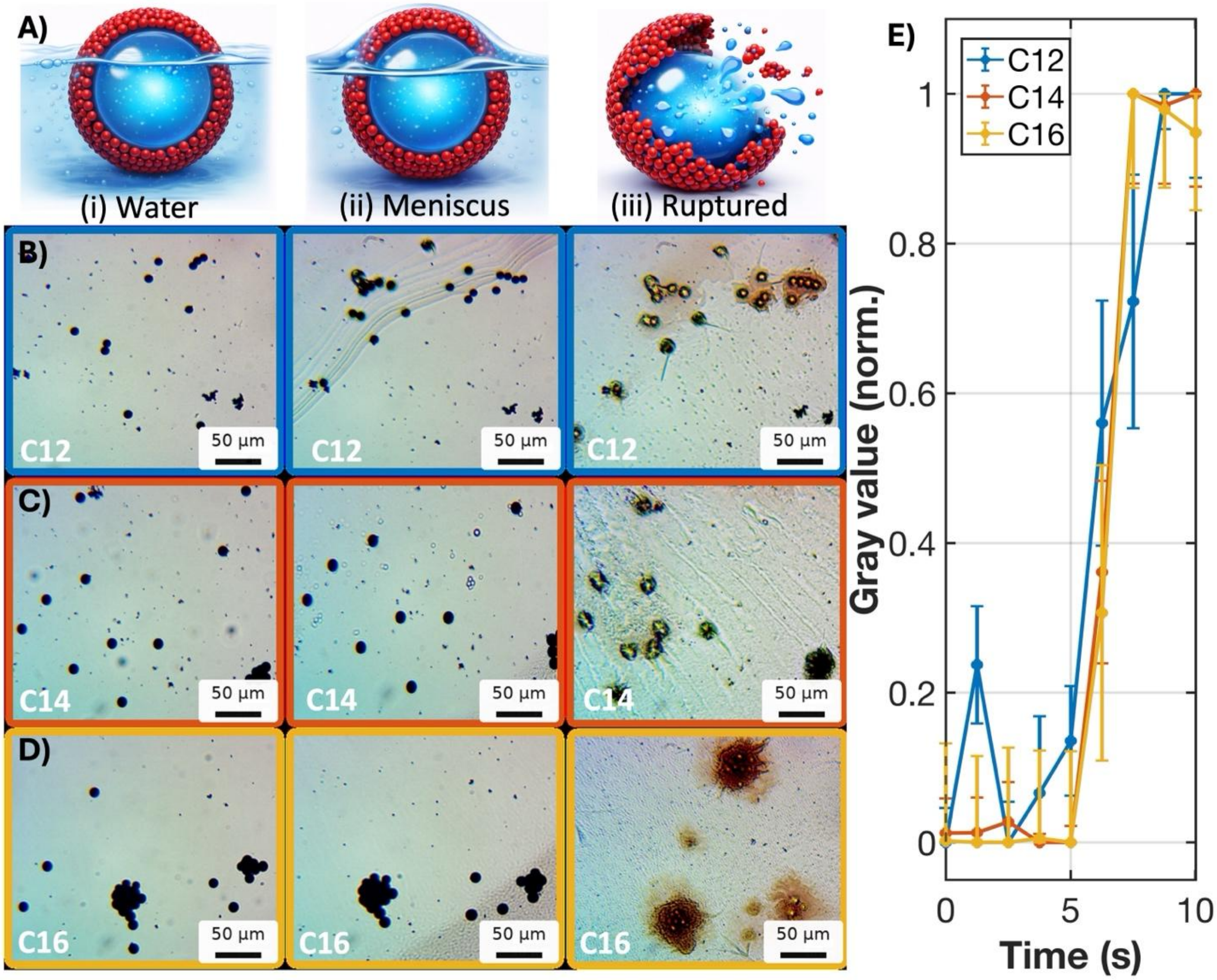

**Figure 5. Meniscus-driven capillary failure of quantum dot (QD) colloidosomes.** (**A**) Conceptual schematic: a receding water–air meniscus traverses a QD colloidosome, where pinning and partial wetting at shell asperities or pores can generate transient capillary stress on the solid shell, promoting rupture and release of core material. (**B–D**) Representative time-lapse optical micrographs of colloidosomes with dodecane (C12), tetradecane (C14), and hexadecane (C16)-rich cores at 0.02% v/v initial alkane loading, shown before, during, and after passage of the drying front. (**E**) Normalized central grayscale intensity versus time for colloidosomes under each alkane condition. For each colloidosome, the grayscale trace was normalized between its pre-front baseline and post-front plateau before averaging across several colloidosomes. Error bars indicate the standard error of the mean across analyzed colloidosomes at each time point.

As water evaporates, the receding water–air interface sweeps across individual colloidosomes, entraining them at the interface. The meniscus may pin locally at asperities or openings in the colloidosome shell rather than retreating smoothly, generating curved air-water interfaces across nanoscale channels. When these openings are at least partially wetted by the aqueous phase, the resulting Laplace pressure imposes a transient capillary stress on the shell. This stress is of the order of the Young-Laplace pressure,[31]

$$\Delta P_{\mathrm{cap}} \approx \frac{2\gamma_{aw} \mid \cos\theta_{eff} \mid}{r_{\mathrm{eff}}},$$

where $\gamma_{aw}$ is the air-water surface tension, $\theta_{eff}$ is an effective receding contact angle of the external air-water meniscus within shell openings, and $r_{eff}$ is an effective channel radius.[32] Using $\gamma_{aw} \approx$ 72-73 mN m$^{-1}$ at 20-25 °C,[33] nanometric channels ($r_{eff}\sim$ 10-100 nm) the resulting capillary stress is on the order of ~ 1-10 MPa for $\mid \cos\theta_{eff} \mid$ of order unity. Such transient stresses are comparable to those reported in nanocapillary wicking systems.[33, 34] We therefore regard meniscus-induced capillary loading as a plausible driving force for shell failure during front passage.

We track the normalized grayscale intensity at the colloidosome center as the drying front passes to quantify the timing and optical signature of this meniscus-triggered rupture and release (Figure 5E; representative image sequences in Figures 5B-D). For each colloidosome, the particle position was re-centered in every frame, and the grayscale value was extracted from a fixed central region of interest. The resulting traces were then normalized and averaged for each alkane condition, with error bars representing the standard error of the mean across the analyzed colloidosomes. Across all three alkanes used, dodecane (C12), tetradecane (C14), and hexadecane (C16), the averaged traces show a rapid rise over approximately 3–6 s, indicating that the response is temporally correlated with front passage. The final plateau level and accompanying halo intensity in the post-front images provide a qualitative proxy for the extent of material release and spreading on the timescale of the experiment. Videos of this process are provided as Supporting Movies 2–4.

**Colloidosome rupture and payload release by a thermal trigger.** The capillary-rupture pathway shown in Figure 5 demonstrates that colloidosomes can undergo mechanical failure via transient meniscus pressures; however, this mechanism is tied to a moving drying front and offers limited temporal and spatial control. Here, we show that temperature can serve as a bulk thermal trigger for colloidosome rupture and payload release, as schematically shown in Figure 6A.

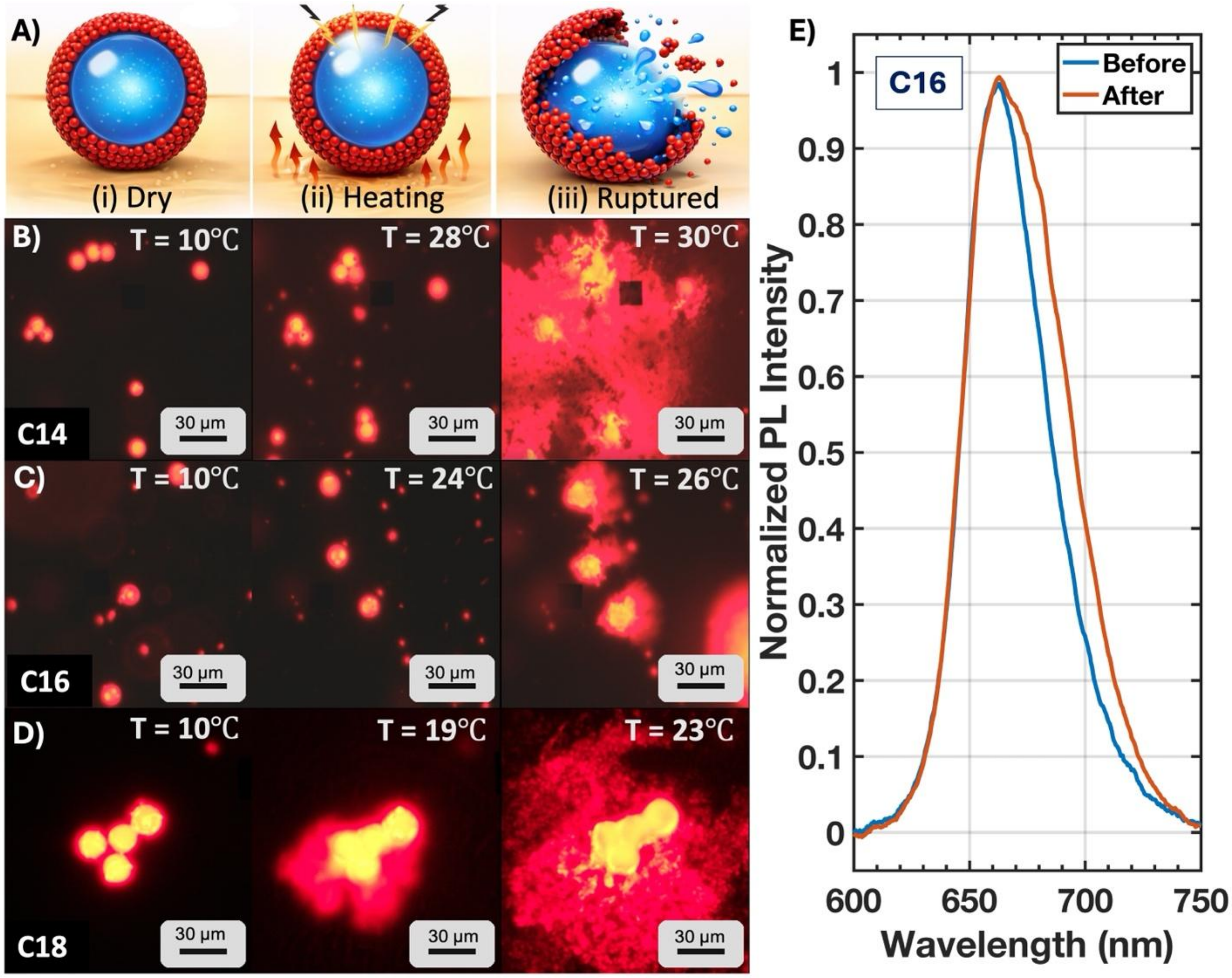

**Figure 6. Thermally activated rupture of liquid-core quantum dot (QD) colloidosomes during warming after cold drying.** (A) Schematic of the thermal pathway: QD colloidosomes are dried while cold to preserve their morphology, then rupture upon warming as core mobility increases and internal stress builds, ejecting QD-containing core material. (B–D) PL micrographs (420 nm excitation) of colloidosomes containing tetradecane (C14, panel B), hexadecane (C16, panel C), and octadecane (C18, panel D) at 0.02 % v/v initial alkane content during a warming sequence (left to right: cold/intact, near-threshold, warm/ruptured), showing abrupt release of QD-containing core material and formation of bright starburst patterns. Corresponding normalized PL spectra for C16 (right) recorded before (blue) and after (red) rupture indicate modest changes to the emission spectrum. Temperature values are indicated in each panel.

We isolate thermal effects from meniscus-driven failure by first cooling colloidosomes to 10 °C on a thermostated substrate and then removing the surrounding aqueous phase under a gentle flow of dry nitrogen. Under these conditions, the colloidosomes remain intact, as shown by the PL images of Figure 6B-D (left panels). The sample was then warmed slowly while imaging, allowing the observation of thermally driven rupture. For each sample (Figure 6B-D), the left image shows the cold, intact state; the middle image captures the intermediate warming regime near the onset of failure; and the right image shows the post-rupture condition, where the emission redistributes into bright "burst" patterns consistent with release and spreading of QDs and core oil.

The rupture temperature decreases with increasing alkane chain length: tetradecane-core colloidosomes rupture near ~30 °C (Figure 6B), whereas hexadecane- and octadecane-core colloidosomes rupture at lower temperatures, ~26 °C and ~23 °C, respectively (Figures 6C-D). We interpret rupture as a mechanically driven failure process, in which constrained thermal expansion of the core and temperature-dependent changes in core mobility impose stress on the jammed QD shell. Rupture is expected when this internal pressure exceeds the shell strength. The lower rupture temperatures observed for longer-chain alkanes are consistent with mechanically driven failure: hexadecane and octadecane have melting/softening transitions closer to the experimental warming range, which may reduce core mobility and promote stress accumulation, whereas tetradecane remains fluid over this range and withstands warming to ~30 °C before rupture.

PL spectra acquired before and after rupture (Figure 6E) show no significant shift in peak position but reveal additional emission on the long-wavelength side. The absence of a spectral shift indicates that rupture primarily redistributes the emitters spatially rather than altering their electronic structure, with QDs acting as optical tracers of colloidosome failure and payload release. The enhanced long-wavelength emission is consistent with the release of QD-containing core material, which may alter the local QD environment and promote energy transfer within the inhomogeneously broadened QD population that can enhance red-side emission, as reported for close-packed CdSe QD solids.[35]

**Colloidosome rupture and payload release by a photothermal trigger.** Thermal loading provides a controlled, global route to colloidosome rupture. Here, we leverage photothermal heating to localize this effect, enabling on-demand rupture of individual colloidosomes while suspended in water. A schematic is shown in Figure 7A.

We use a focused near-infrared (NIR) laser to trap and monitor individual colloidosomes. A single colloidosome suspended in an aqueous SDS solution is held in place with a focused 976 nm optical trap generated using opposing 100× objectives (NA 1.25); a detailed schematic of the trapping/detection geometry and spot-size estimates are provided in the Supporting Information. At 50 mW of focused laser power, the trapped colloidosome appears intact and unchanged, whereas at 100 mW it remains trapped while a faint red-orange QD emission becomes visible (Figure 7B). Figure S2 shows that 976 nm lies below the absorption edge of CdSe/CdS QDs, therefore the observed red-orange emission is unlikely to arise from direct band-edge excitation. Power-dependent measurements under continuous-wave 976 nm excitation show a near-quadratic dependence of the integrated PL intensity on excitation power, with a fitted log-log slope of 2.00 (Figure S8). This suggests two-photon excitation as the excitation mechanism. Thus, the 976 nm trap enables two-photon-induced spectral interrogation of a single QD colloidosome.

We measure the two-photon emission of a single QD colloidosome while varying the collection position. When the detection spot is centered on the colloidosome, the PL spectrum is dominated by the broad QD emission envelope, as shown in Figure 7C. However, shifting the detection spot toward the colloidosome periphery reveals narrow resonant features that emerge from the broad emission envelope and intensify relative to the background. This behavior is consistent with the coupling of light emitted by QDs with WGMs of the colloidosome. These photonic modes allow light to propagate along the surface of the colloidosome; collecting near the edge therefore enhances the cavity-mediated contribution relative to the bulk-like, non-resonant QD emission collected near the particle center.

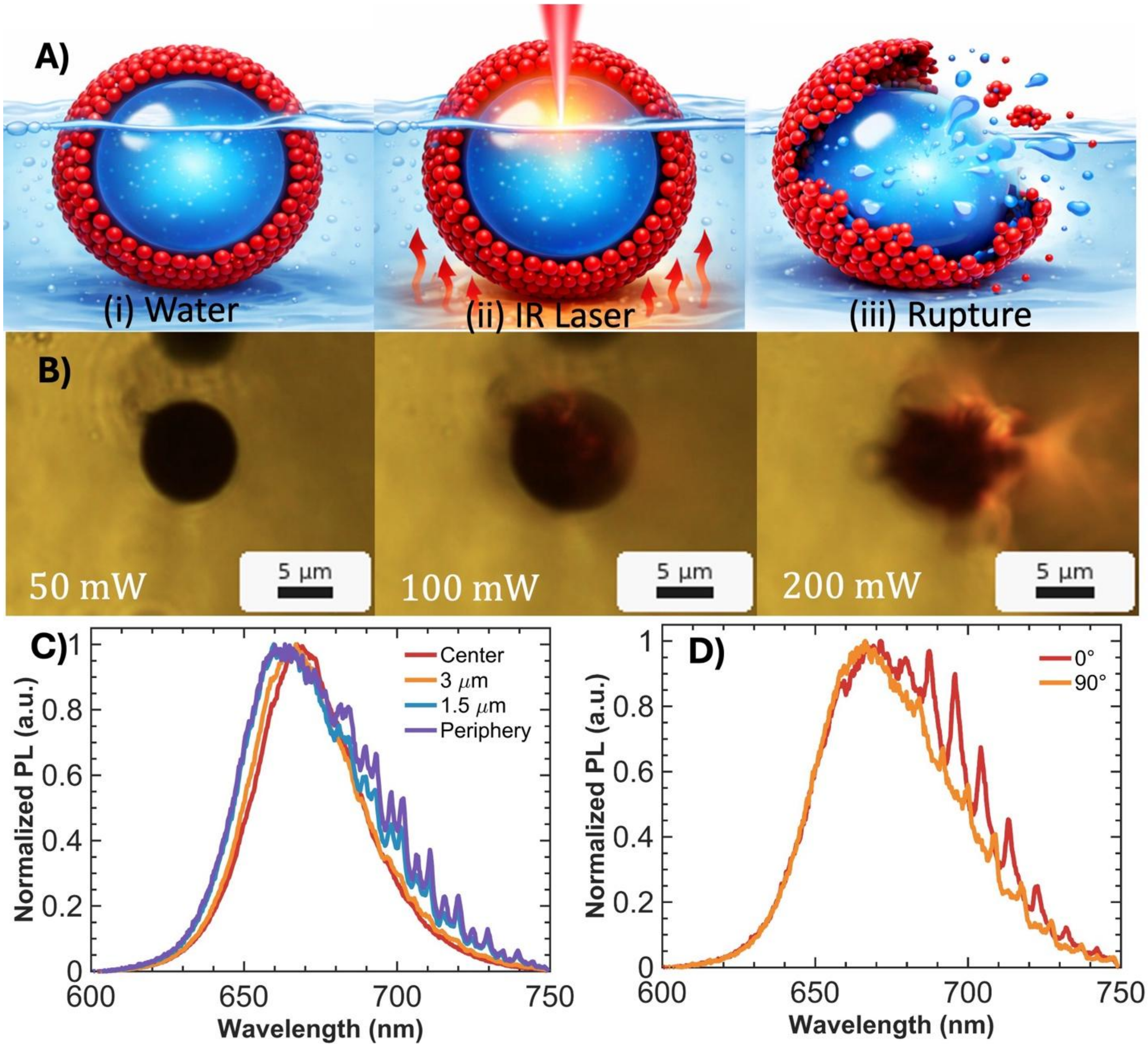

**Figure 7. Local rupture of a liquid-core quantum dot (QD) colloidosome by focused near-infrared (NIR) excitation**. A) Schematic of NIR-triggered photothermal rupture of a QD

colloidosome in water. (B) Optical micrographs of a single colloidosome under increasing 976 nm excitation power (50, 100, 200 mW), showing intact, emissive, and ruptured states, respectively. (C) Normalized PL spectra collected from different lateral collection positions, showing stronger narrow resonant features toward the shell periphery, consistent with whispering-gallery modes. (D) Polarization-resolved PL spectra acquired near the periphery of the colloidosome, showing angle-dependent changes in resonant mode intensities.

Polarization-resolved spectra collected near the periphery of the colloidosome using a rotating linear polarizer show that the relative intensities of the narrow resonant modes change with analyzer angle, Figure 7D. As the analyzer rotates, several modes strengthen while others weaken, indicating that the cavity emission contains polarization-dependent components. This behavior is consistent with polarization-dependent contributions from WGM resonances with different orientations, including transverse electric (TE)- and transverse magnetic (TM)-like character. Rotating the analyzer projects different mixtures of these cavity modes into the detection path.

Additional evidence comes from the analysis of spectral separation between adjacent resonances. The two adjacent spacings among the high-fluence peaks in Figure 4A are 4.04 and 4.47 nm, while the passive NIR-excited modes fitted in Figure S9 exhibit adjacent spacings of 3.5-5.0 nm, with a mean spacing of approximately 4.4 nm. Although these spectra were acquired from different individual colloidosomes under different excitation conditions, the agreement in characteristic modal spacing supports a common WGM-cavity origin. Taken together, the clear log–log slope change near 2.8 mJ/cm$^2$ (Figure S7), strong spectral narrowing, discrete cavity-defined peaks, peripheral enhancement of the resonant features, polarization-dependent mode intensities, passive-cavity resonances, loss of the narrow features following shell rupture, morphology controls, and numerical eigenmode calculations support assignment of the high-fluence response as WGM-assisted lasing. Neither ASE alone nor passive cavity-modulated photoluminescence alone accounts for this combined set of observations.

Increasing the NIR laser power to 200 mW causes the colloidosome to rupture, as shown in Figure 7B. The rupture is accompanied by a brief flash of QD emission, consistent with sudden local heating followed by pressure release within the confined colloidosome volume. The jammed shell ruptures and the liquid core is released, dispersing QD-containing material into the surrounding aqueous medium (see Supporting movie 5). Under focused 976 nm irradiation, local heating near the trapped colloidosome can arise from absorption by both the QD-rich shell and the surrounding aqueous medium. We therefore attribute rupture primarily to photothermal loading under sustained trapping rather than to direct mechanical fracture by the optical field, see section 2.0 in SI for order-of-magnitude analysis. These results show that individual suspended QD colloidosomes can be spectroscopically interrogated and ruptured on demand by focused NIR light. This enables localized payload release without the need for global drying or uniform heating. Such NIR-addressable activation could support spatially patterned marking, localized sensing, and on-demand release in microscale fluidic and photonic environments.

## Conclusion

We introduce QD colloidosomes as self-assembled microcavities in which optical feedback, material efficiency, and triggered release are controlled by shell architecture. By varying alkane content during assembly, we access a morphological continuum spanning solid supraparticles, liquid-core/solid-shell colloidosomes, and microporous hollow structures. This continuum identifies shell continuity, rather than the presence of a solid core, as the critical structural requirement for WGM feedback and lasing. Intact colloidosomes sustain lasing with thresholds and quality factors comparable to solid supraparticles while requiring substantially less QD material, whereas porous shells disrupt optical feedback and remain broadband even at high pump fluence. The same shell architecture that supports lasing also provides a built-in failure pathway: capillary stresses, bulk heating, and localized NIR excitation can drive irreversible rupture of intact colloidosomes, extinguishing cavity-defined emission and releasing QD-containing core material. These results establish QD colloidosomes as switchable photonic microcontainers and define a design strategy in which shell integrity serves as the common control parameter for light generation, optical shutdown, and triggered release in self-assembled semiconductor microcavities.

# Supporting Information

for

# Quantum Dot Colloidosomes as Triggerable Microlasers

Cristian Gonzalez[1], Saranya Subramanian,[4] Marco Reale,[4,6] Giuseppe Soligno,[5] Ilia Geints[3], Siyuan Yin[1], Claire Y. Kang[1], Ricardo Ronquillo,[1] Gary Chen[1], Marco Cannas,[4,6] Cherie R. Kagan[1,2,3], Alice Sciortino[4], Michael Engel,[5] Fabrizio Messina[4], Christopher B. Murray[1,2]*, Emanuele Marino[4]*

[1] Department of Chemistry, [2] Department of Materials Science & Engineering, and [3] Department of Electrical & Systems Engineering, University of Pennsylvania, Philadelphia (PA), USA.

[4] Department of Physics and Chemistry – Emilio Segrè, University of Palermo, Palermo, Italy

[5]Institute for Multiscale Simulation, IZNF, Friedrich-Alexander-Universität Erlangen-Nürnberg, 91058 Erlangen, Germany.

[6]INSTM – National Interuniversity Consortium of Materials Science and Technology, Florence, Italy

*Corresponding author: emanuele.marino@unipa.it, cbmurray@sas.upenn.edu.

# Methods:

## Synthesis of CdSe Core Quantum Dots

CdSe core quantum dots used in this work were synthesized according to previously reported literature procedures.[1, 2] The purified CdSe core dispersion was stored in hexane under an inert atmosphere prior to CdS shell growth.

## Growth of 4-Monolayer CdS shells

CdSe/CdS core/shell quantum dots with approximately four monolayers of CdS were prepared by shelling the CdSe cores according to previously reported literature procedures.[2, 3] The purified CdSe/CdS core/shell quantum dots were dispersed in toluene and stored under an inert atmosphere prior to use.

## Electron Microscopy

Scanning electron microscopy (SEM) was performed on a Tescan S8252X using an Everhart–Thornley (E–T) secondary electron detector. Imaging was carried out in UH-resolution scan mode at a working distance of 5.82 mm and a beam current of 100 pA. The accelerating voltage was adjusted as needed to optimize contrast of surface morphology; unless otherwise noted, representative micrographs were acquired at 10 keV. The stage tilt was maintained at 0° for all SEM images.

Transmission electron microscopy (TEM) was performed on a JEOL-1400 operated at 120 kV. TEM samples were prepared by drop-casting dilute dispersions onto carbon-film-coated 300-mesh copper grids, followed by solvent evaporation under ambient conditions prior to imaging.

## Microfluidic Source-Sink Emulsion

Monodisperse nanocrystal-containing toluene droplets were generated using a commercial glass droplet-generator chip (Darwin Microfluidics, T-26) connected to a multichannel pressure regulator (Elveflow, OB1MK3+). The nanocrystal dispersion in toluene, aqueous continuous phase, and hexadecane-in-water sink emulsion were introduced at pressures of 1, 1, and 2 bar, respectively. The outlet was connected to 11.5 m of PFA tubing with an inner diameter of 0.5 mm, wrapped around a cylindrical copper rod equipped with a heating cartridge and thermocouple and maintained at 65 °C. During the approximately 4 min residence time in the heated tubing, toluene transferred from the nanocrystal-containing source droplets into the nanoscale hexadecane sink droplets, causing the source droplets to shrink and the nanocrystals to densify into supraparticles. The resulting dispersion was collected in 20 mL scintillation vials containing 5 mL of sink emulsion. The collection vials were subsequently left uncapped and undisturbed on a hot plate set to 50 °C for 24 h to remove residual toluene.

## Numerical Electromagnetic Simulations

Finite-element eigenmode calculations were performed in COMSOL Multiphysics using the Wave Optics Module and a two-dimensional axisymmetric formulation. The idealized colloidosome was represented by concentric spherical regions with a fixed outer radius of 4.20 μm: a lossless alkane core with refractive index $n = 1.430$, an effective CdSe/CdS QD shell with $n = 2.184$, and a surrounding aqueous medium with $n = 1.331$. The shell thickness was set to 250, 300, or 350 nm while the outer radius and material parameters were held fixed. The surrounding water domain was terminated by a perfectly matched layer. Eigenfrequency calculations were performed using azimuthal mode number $m = 72$, and the same TE-like $(m,l) = (72,72)$ mode branch was tracked across the three thicknesses. The quality factor was calculated from the complex eigenfrequency as $Q = \mathrm{Re}(f)/(2|\mathrm{Im}(f)|)$.

Complementary three-dimensional finite-difference time-domain calculations were performed using Tidy3D (version 2.12.0) with the same concentric-sphere geometry and lossless material parameters. Shell thickness was varied from 25 to 350 nm by changing the inner radius according to $R_{inner} = R_{outer} - t$ while holding the 4.20 μm outer radius fixed. The computational domain extended 1.0 μm into water beyond the resonator and was terminated by 12-cell perfectly matched layers on all sides. A broadband z-directed current source proportional to $\cos(72\varphi)$, positioned within the equatorial shell, excited the $m = 72$ mode over 580–760 nm. Symmetry conditions (1,1,−1) were applied. The time-dependent field decay was recorded at four positions around the shell and analyzed using the Tidy3D ResonanceFinder to obtain the resonance frequency and radiative Q. The shell-localization fraction was calculated as the electric-field intensity integrated within the QD shell divided by that integrated over the non-PML computational domain. Because the materials were treated as lossless and the shells as smooth and continuous, the calculated Q values include radiative and numerical leakage but exclude material absorption, gain, roughness, porosity, and QD-packing disorder.

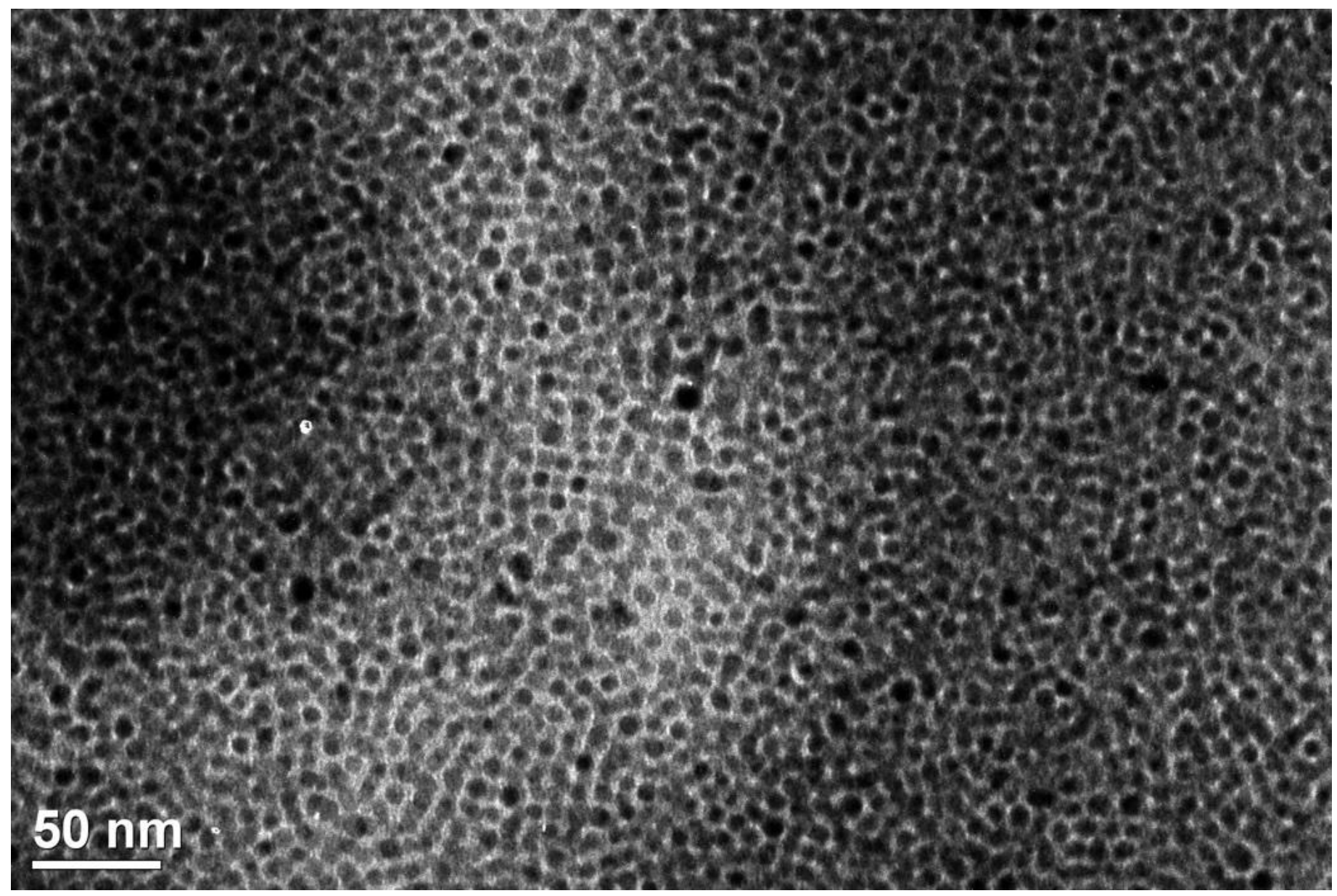

**Figure S1:** Transmission electron micrograph of the CdSe/CdS sample drop-casted on a Cu-Carbon FILM 300 mesh grid.

## Optical Measurements:

## Fluence Dependent Lasing

Emission spectra for fluence-dependent measurements were acquired using a home-built optical set-up. As an optical pump source, a Q-switched Nd:YAG laser (Vibrant, OPOTEK) was used, whose third harmonic pumps an optical parametric oscillator, producing tunable pulses with a FWHM duration of ≈2 ns at a repetition rate of 10 Hz. For the fluence-dependent experiments, the excitation wavelength was set to 500 nm. The pump beam was focused onto the sample using a lens in a wide-field reflection geometry. To selectively collect the emission from individual suprastructures, an appropriate combination of a 40X/0.65NA objective with an optical fiber was used. The emitted light was filtered through a 580 nm long-pass filter and directed to a spectrophotometer consisting of a monochromator (Acton, SpectraPro 2300i), equipped with a 300 grooves mm−1 grating, providing a spectral resolution of 1 nm and an intensified CCD camera (Princeton Instruments, Pixis 400). For measurements on intact (or hollow) colloidosomes, we created a 35 mm × 15 mm rectangular teflon mask placed on a microscope glass slide to form a sealed chamber. First, 20 µL of deionized water was introduced into the chamber, followed by 10 µL of the intact (or hollow) colloidosome dispersion, and the chamber was immediately sealed with a cover glass. Measurements were carried out quickly to minimize water evaporation.

## Optical Tweezing

Photoluminescence measurements and photothermal rupture experiments were performed on a home-built optical tweezers micro-PL platform. Samples were loaded into a gasketed chamber assembled by placing a 0.03″-thick adhesive silicone gasket on a glass microscope slide, filling the well with 50 μL of colloidosomes dispersed in water containing surfactant (SDS), and sealing with a No. 1.5 cover glass. Measurements were performed on an inverted microscope with the cover glass facing the objective(s). Individual colloidosomes were trapped using a focused near-infrared laser (λ ≈ 976 nm) in a counter-propagating dual-objective geometry (two opposing 100×, NA 1.25 oil-immersion objectives). Trap power at the sample was varied (50–200 mW) to induce controlled photothermal loading and, at high power, rupture. Emission from a trapped colloidosome was collected through the microscope optics, spectrally filtered to reject the trapping laser, and coupled into a multimode optical fiber feeding a Horiba iHR550 spectrometer (150 g $mm^{-1}$ grating). The spectrometer entrance slit width was set to 100 μm. Under these conditions, the spectral resolution was primarily limited by the entrance slit width and was estimated to be $\Delta\lambda$ =1.2 nm based on the spectrometer dispersion, in agreement with the experimentally measured value of $\Delta\lambda$ ≈1.2 nm. For polarization-resolved measurements, a linear polarizer mounted on a rotation stage was placed in the collection path prior to fiber coupling.

Peak incident intensity (power density) was estimated assuming a Gaussian beam profile. Focal-spot FWHM diameter was measured to be approximately 1.1 $\mu m$, the peak intensity (power density) was calculated as

$$I_0 = \frac{4P\ln(2)}{\pi d_{\mathrm{FWHM}}^2}$$

For example, $I_0 \approx 7.3\ \mathrm{MW/cm^2}$ at 100 mW average incident power.
(actual local intensity experienced by some of the QDs may be different and hard to quantify because the micron-scale semi-transparent superparticle can modify the internal optical field through refraction, scattering, and internal field redistribution).

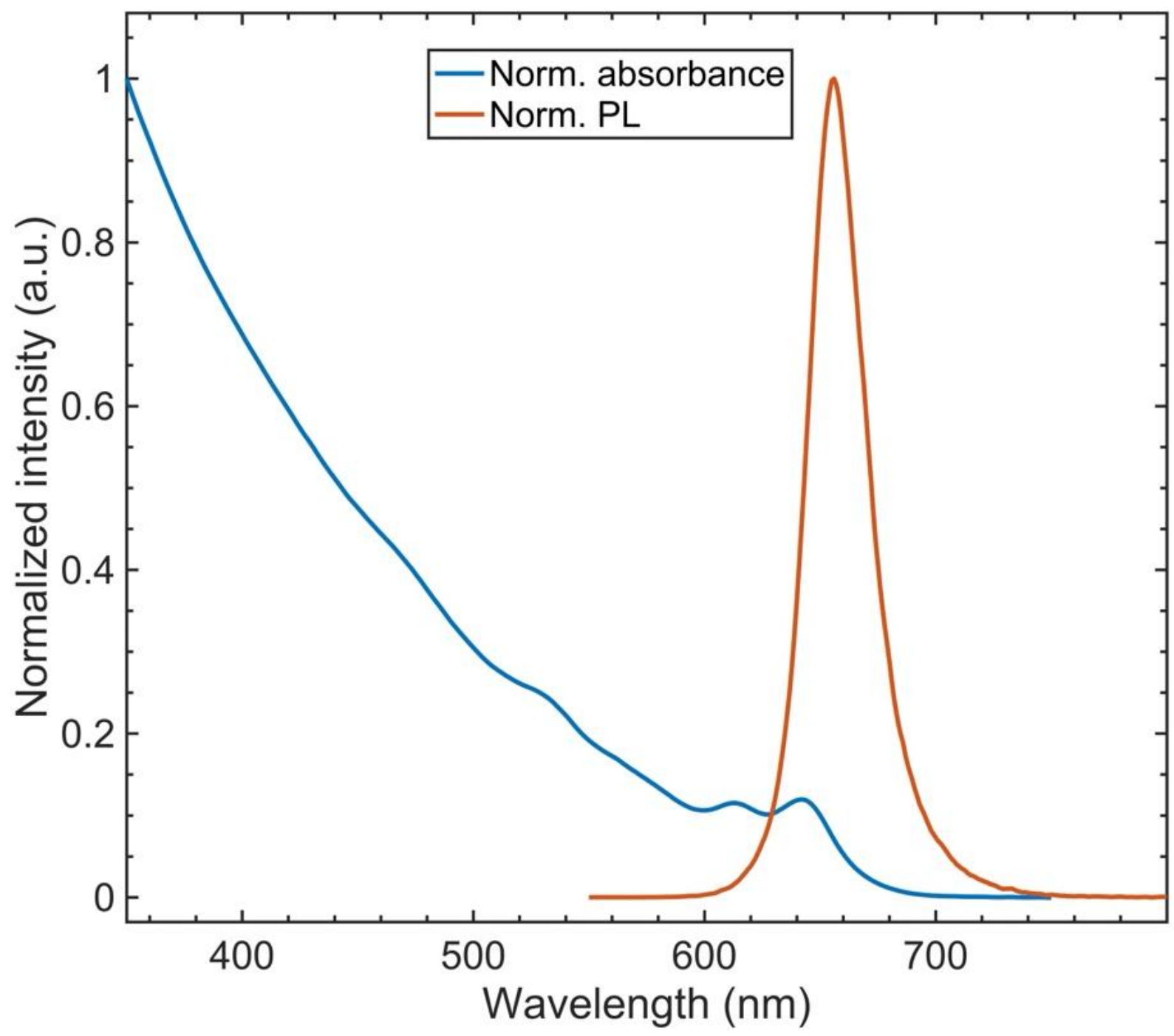

**Figure S2:** Absorption and photoluminescence (PL) of Core/shell CdSe/CdS quantum dots (QDs)

## Section 1.0: Adsorption Energy of Spherical and Cubic Particles at the Oil–Water Interface

Here we illustrate how we calculated the adsorption energy $E$ [Eq. 1]. Firstly, for convenience, we exploit Young's Law [Eq.2] to rewrite $E$ as it follows

$$E = \gamma_{ow}(S - S_0 + \cos\theta\, W_w) \quad . \qquad \text{Eq.1}$$

At the equilibrium, the spherical particle sits at the flat oil-water interface forming a contact angle $\theta$ in the oil phase, see sketch in Fig. S3. Therefore, $E$ can be computed analytically for the spherical particle using that $S - S_0 = -\pi(R\sin\theta)^2$ and $W_w = 2\pi R^2(1 - \cos\theta)$, obtaining

$$E = -\gamma_{ow}\,\pi\,R^2\,(1 - \cos\theta)^2.$$

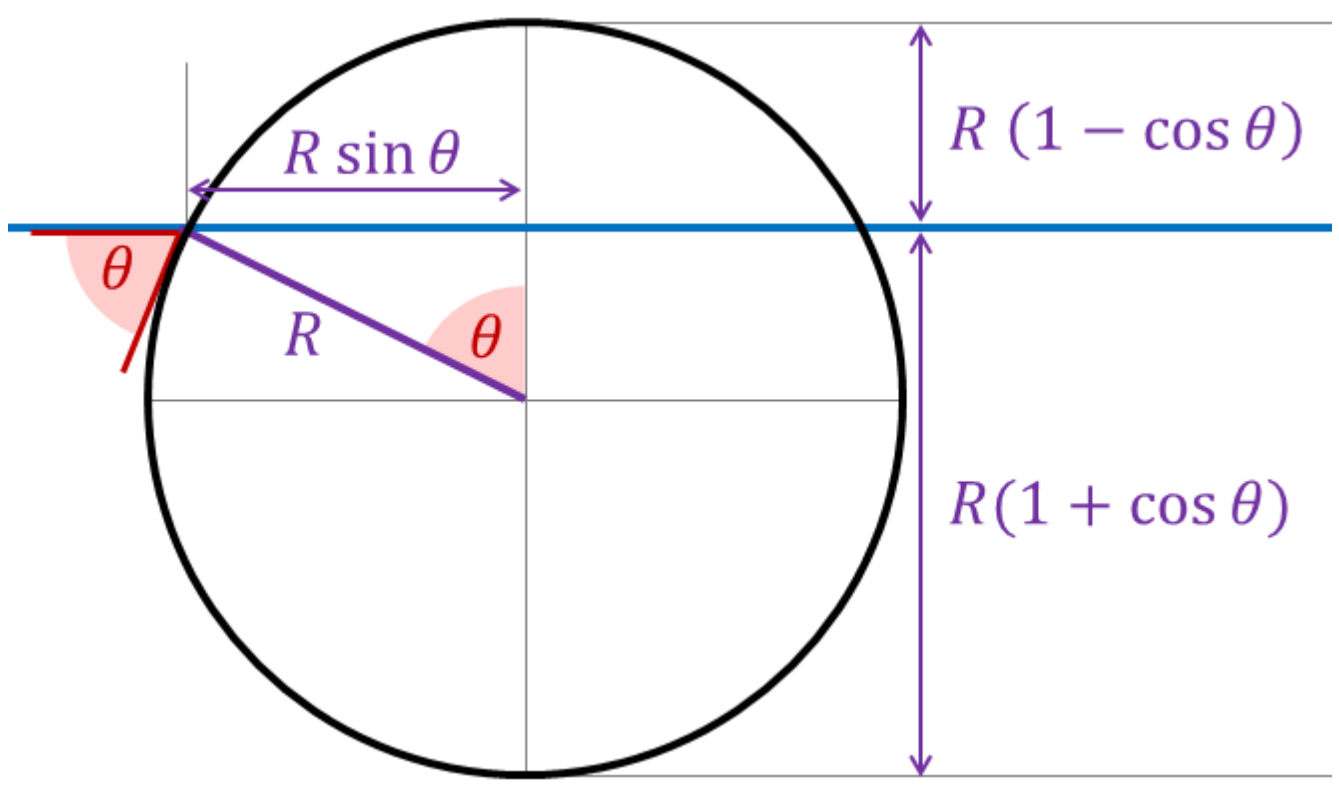

**Figure S3:** Sketch of a spherical particle adsorbed at its equilibrium height at a flat fluid-fluid interface, where $\theta$ is Young's contact angle and $R$ is the sphere's radius.

For the cubic particle, we calculate $E$, minimized over the cube's height at the interface plane, with respect to the cube's orientation $\varphi$, $\psi$ at the interface plane, see Fig. S3. Then, at the equilibrium, $E$ has the value corresponding to the cube's orientation with minimum $E$, i.e. when the cube has one face parallel to the interface and wet by the water phase, see inset "a" in Fig. S3.

To obtain $E(\varphi, \psi)$, we simulate the equilibrium shape of the fluid-fluid interface near the cube, for each considered cube's orientation, and then we compute $S$ and $W_w$. These calculations are performed using Interface Equilibrator.[4] The initial shape of the fluid-fluid interface mesh is a flat mesh with a hole in its center, by where the cube is placed in the desired orientation. Then, the boundary of such a hole is constrained on the cube, while the mesh's outer boundary is constrained on a vertical wall enclosing the system and with Young's contact angle 90° (to reproduce a flat interface far away from the particle). Then, the simulation of the fluid-fluid interface's equilibrium shape is performed with the option of not constraining the fluids' volumes, so that the fluid-fluid interface's equilibrium shape is obtained for the equilibrium height of the cube at the interfacial plane. Finally, $S$ and $W_w$ are obtained from the fluid-fluid interface's equilibrium shape using Interface Equilibrator built-in functions.

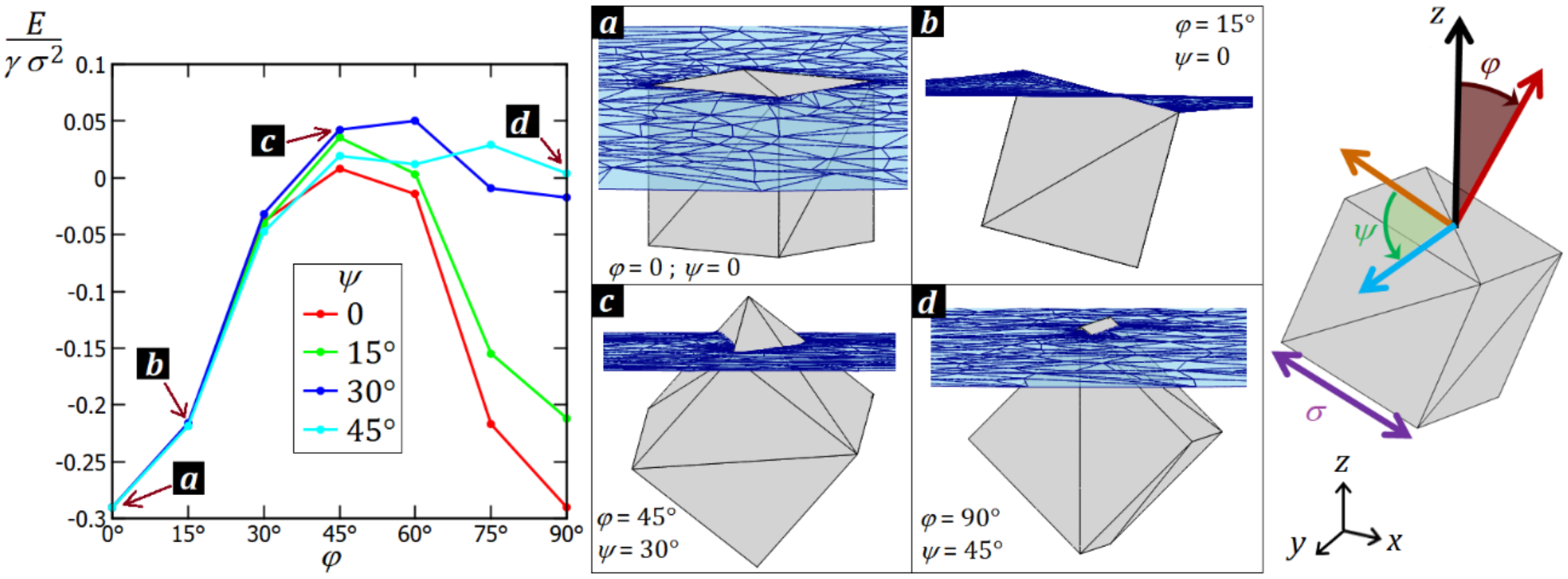

**Figure S4:** Adsorption energy $E$ [Eq.1] of a cubic particle with side $\sigma$ at a fluid-fluid interface, minimized over the cube's height at the interface plane (parallel to $z = 0$) and with respect to the cube's orientation $\varphi$, $\psi$, where $\varphi$ is the angle formed by the cube's vertical axis with the $z$ axis

and $\psi$ is the internal Euler angle around the cube's axis (see sketch on the right). The fluid-fluid interfacial tension is $\gamma$ and $\cos\theta = 0.7$, with $\theta$ the Young's contact angle. The insets (a-d) show a 3D view, near the cube, of the fluid-fluid interface's equilibrium shape computed with Interface Equilibrator,[4] for various orientations of the cube at the interface.

## Section 2.0: Order-of-Magnitude Estimate for Photothermal vs Optical-Force Loading During NIR Trapping

Here, we provide an order-of-magnitude estimate to assess whether colloidosome rupture under sustained 976 nm trapping is plausibly driven by photothermal loading (local heating and pressure/cavitation events) rather than direct mechanical fracture by optical forces.

### Heat diffusion timescale in water

A characteristic timescale for thermal diffusion over a length scale $L$is

$$t_{\mathrm{diff}} \sim \frac{L^2}{\alpha},$$

where $\alpha$is the thermal diffusivity of the surrounding medium. For water near room temperature, $\alpha \sim 10^{-7}$ m$^2$ s$^{-1}$.[5] Using $L \sim 1\text{–}10\mu$m , representative of micron-scale gradients around a trapped colloidosome,

$$t_{\mathrm{diff}}(1\ \mu\mathrm{m}) \sim \frac{(10^{-6}\ \mathrm{m})^2}{10^{-7}\ \mathrm{m}^2\ \mathrm{s}^{-1}} \sim 10^{-5}\ \mathrm{s}, \qquad t_{\mathrm{diff}}(10\mu\mathrm{m}) \sim \frac{(10^{-5}\ \mathrm{m})^2}{10^{-7}\ \mathrm{m}^2\ \mathrm{s}^{-1}} \sim 10^{-3}\ \mathrm{s}.$$

Thus, heat diffuses across micron length scales on microsecond–millisecond timescales, far shorter than the seconds-long trapping dwell time before rupture. The temperature field, therefore, rapidly approaches a quasi-steady profile under continuous NIR irradiation.

### Steady-state temperature rise near a localized heat source

If absorbed optical power $P_{\mathrm{abs}}$is converted to heat and dissipated into water, the steady-state temperature rise at distance $r$from an effectively localized heat source in an infinite medium scales as

$$\Delta T(r) \sim \frac{P_{\mathrm{abs}}}{4\pi k r},$$

where $k$is the thermal conductivity of the surrounding medium (water: $k \approx 0.6\ \mathrm{W\,m^{-1}\,K^{-1}}$).[5] The factor $4\pi r$ arises from radial heat flow through spherical surfaces ($4\pi r^2$) combined with the $1/r$ steady-state temperature profile around a point-like source.

We relate absorbed to incident trap power via

$$P_{\mathrm{abs}} = \eta P,$$

where $P$is the incident trap power at the sample, and $\eta$is an effective absorption fraction (lumping absorption by the QD shell and any other absorbing components). For the rupture condition $P \approx 200$ mW, $P_{\mathrm{abs}} \approx 0.2\,\eta$ W. Evaluating at $r \sim 5\mu$m (order of the colloidosome radius),

$$\Delta T(r \sim 5\ \mu\mathrm{m}) \sim \frac{0.2\,\eta}{4\pi(0.6)(5 \times 10^{-6})} \approx 5.3 \times 10^{3}\,\eta\ \mathrm{K}.$$

For modest absorption fractions:

- $\eta = 10^{-3}$(0.1% absorbed): $\Delta T \sim 5$ K
- $\eta = 5 \times 10^{-3}$(0.5% absorbed): $\Delta T \sim 25$ K
- $\eta = 10^{-2}$(1% absorbed): $\Delta T \sim 50$ K

at micron distances. Temperature rises of tens of kelvin are sufficient to substantially increase vapor pressure and promote cavitation-like failure in a confined volatile liquid core, consistent with the observed abrupt rupture and transient emission flash.

*Note:* This estimate intentionally captures scaling. Interfaces (glass proximity), finite source size, and convection modify prefactors but do not change the core conclusion that modest absorbed power can generate substantial local heating at micron length scales.

**Optical-force scale**

Optical tweezers typically apply forces in the piconewton range (order 1–100 pN, depending on trap stiffness and object properties).[6] A conservative pressure scale associated with a pN-level force distributed over a micron-scale area $A \sim \pi R^2$with $R \sim 5\ \mu$m is

$$p \sim \frac{F}{A} \sim \frac{10^{-11}\text{–}10^{-10}}{\pi(5 \times 10^{-6})^2} \approx 0.1\text{–}1\ \mathrm{Pa}.$$

This pressure scale is extremely small compared with stresses typically associated with fracture of a jammed particulate shell and, importantly, compared with the transient stresses expected during thermally driven pressure release/cavitation within a confined liquid volume. Moreover, in a counter-propagating trapping geometry, net scattering forces are reduced, further decreasing the likelihood of rupture driven by asymmetric radiation pressure.

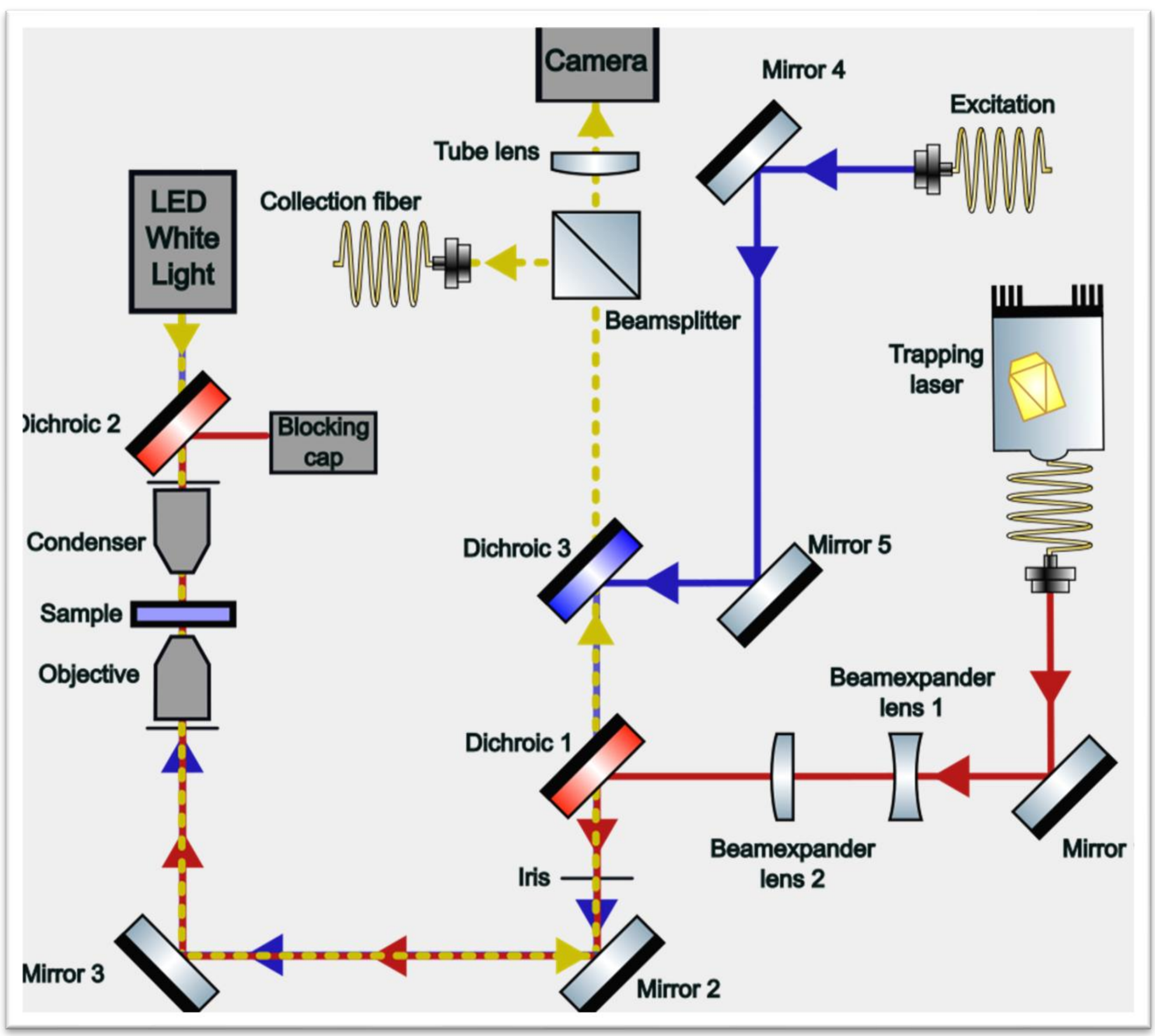

**Figure S5:** Optical layout for combined optical tweezing, two-photon excitation, imaging, and spectral collection. The trapping laser is expanded and coupled into the objective through dichroic 1, while the excitation beam is directed by mirrors 4 and 5 and combined through dichroic 3. White-light illumination from the LED is transmitted through the condenser for sample visualization, with a blocking cap used to suppress unwanted illumination. Emission from the trapped liquid-laser sample is collected through the objective and routed by the beamsplitter either to the camera for imaging or to the collection fiber for spectral analysis.

## Section 3.0: Active Quality Factor Analysis

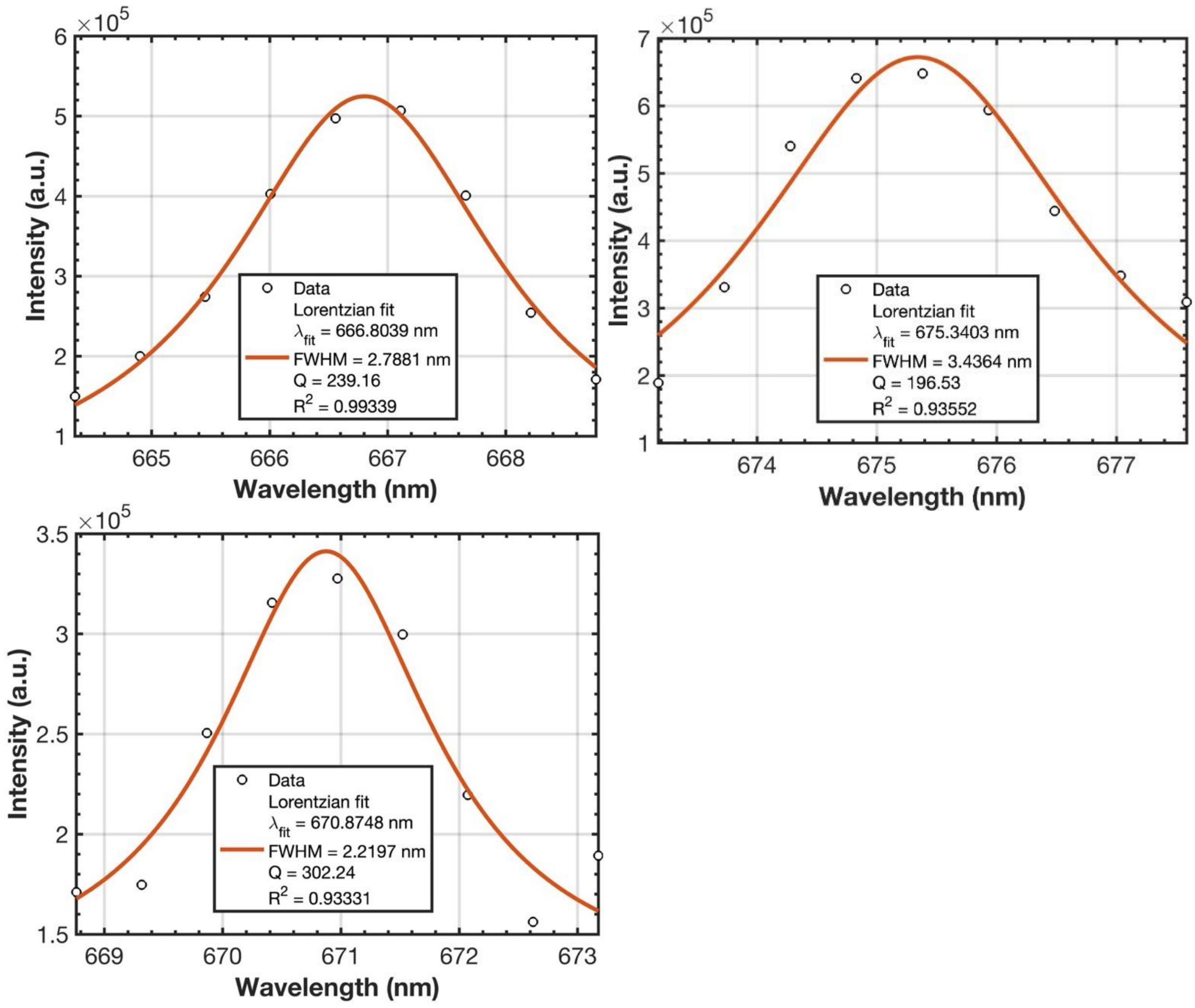

**Figure S6:** Lorentzian fitting of individual lasing modes from an intact QD colloidosome. Lorentzian fits to the three dominant narrow emission modes measured from an intact QD colloidosome at the highest pump fluence, 14.0 mJ cm$^{-2}$. Black open circles represent the experimental spectra, and orange solid lines represent Lorentzian fits used to extract the peak center wavelength, full width at half maximum (FWHM), and active quality factor $Q = \lambda/\Delta\lambda$. The fitted modes are centered at 666.83, 670.87, and 675.34 nm, with FWHM values of 2.79, 2.22, and 3.45 nm, corresponding to active Q factors of 239, 302, and 196, respectively. These fits support the assignment of the narrow spectral features to cavity-defined lasing modes from the intact colloidosome.

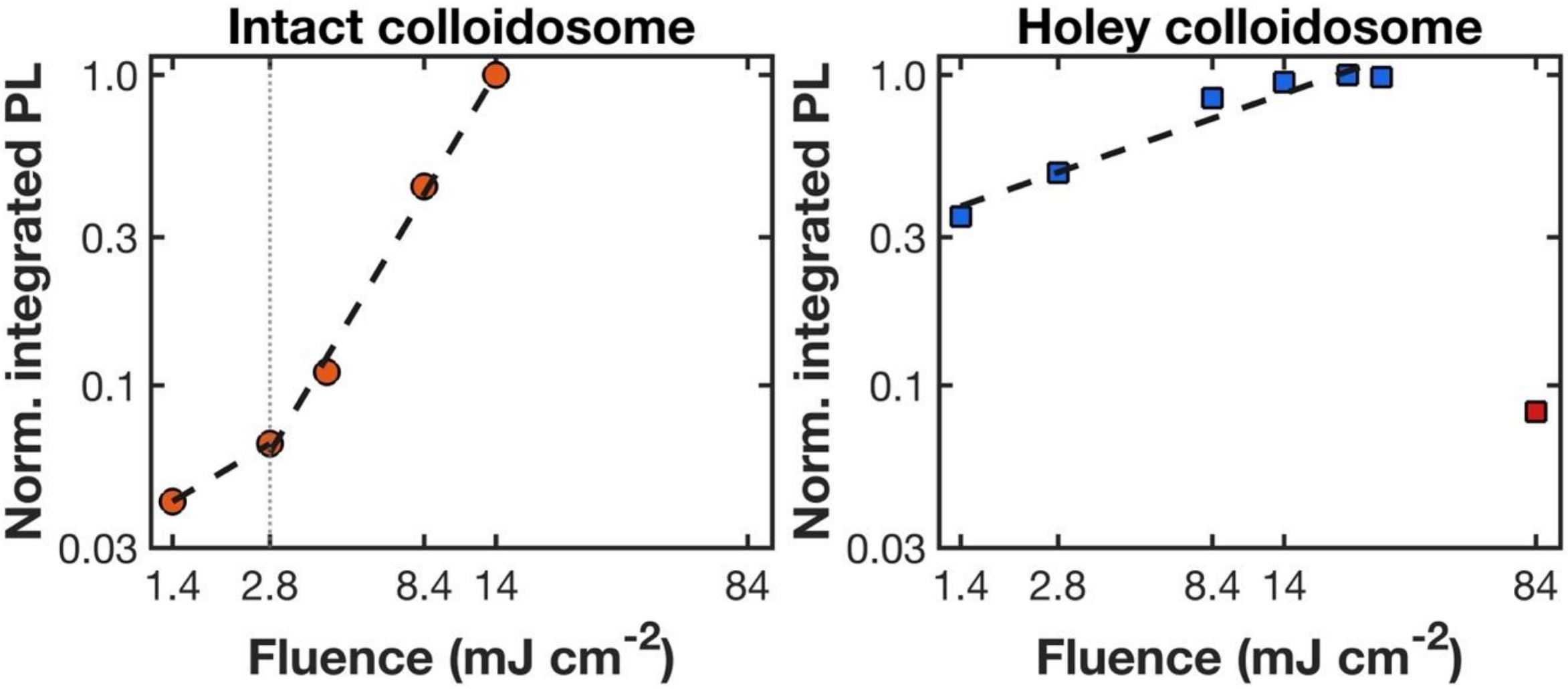

**Figure S7:** Threshold analysis of fluence-dependent emission from intact and holey QD colloidosomes. Normalized integrated PL intensity plotted as a function of pump fluence for an intact QD colloidosome and a holey/porous-shell colloidosome. Both panels are shown on log–log axes. The intact colloidosome exhibits a clear slope change near 2.8 mJ/cm$^2$, coincident with the emergence of ASE-like spectral features and subsequent cavity-mode lasing. In contrast, the holey colloidosome shows a gradual increase in broadband emission without a clear threshold-like knee, followed by an intensity drop at 84 mJ/cm$^2$ consistent with high-fluence photodamage or structural disintegration. Dashed lines are guides to the eye highlighting the low- and high-fluence emission trends.

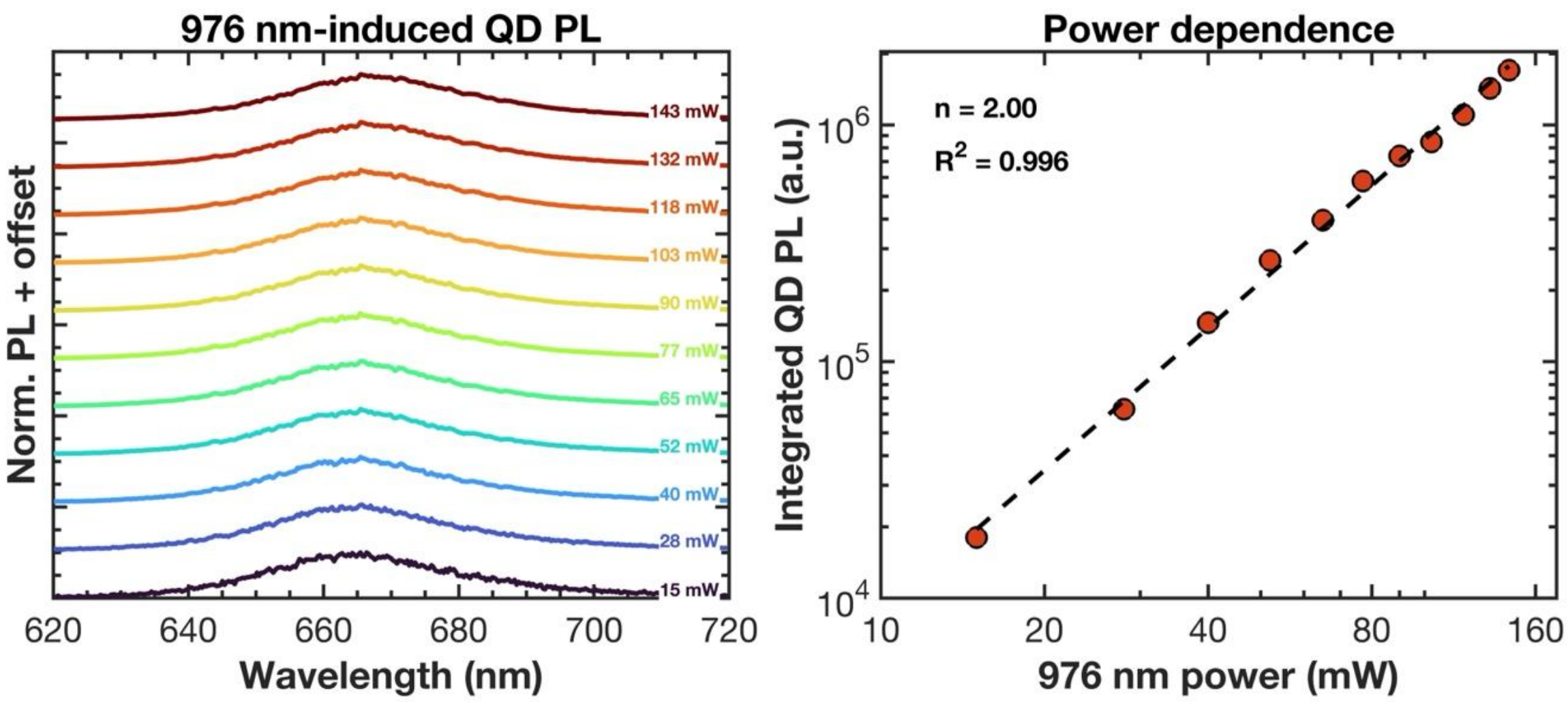

**Figure S8:** Stacked photoluminescence spectra collected from a QD colloidosome under continuous-wave 976 nm excitation with increasing calibrated excitation power. The spectra show the emergence and growth of emission in the QD PL band, consistent with NIR-induced excitation of the CdSe/CdS QDs. (B) Integrated QD PL intensity plotted as a function of calibrated 976 nm excitation power on log–log axes. The data follow a power-law dependence, $I_{PL} \propto P^n$, with a near-quadratic exponent, supporting assignment of the 976 nm-induced emission to two-photon-excited QD photoluminescence.

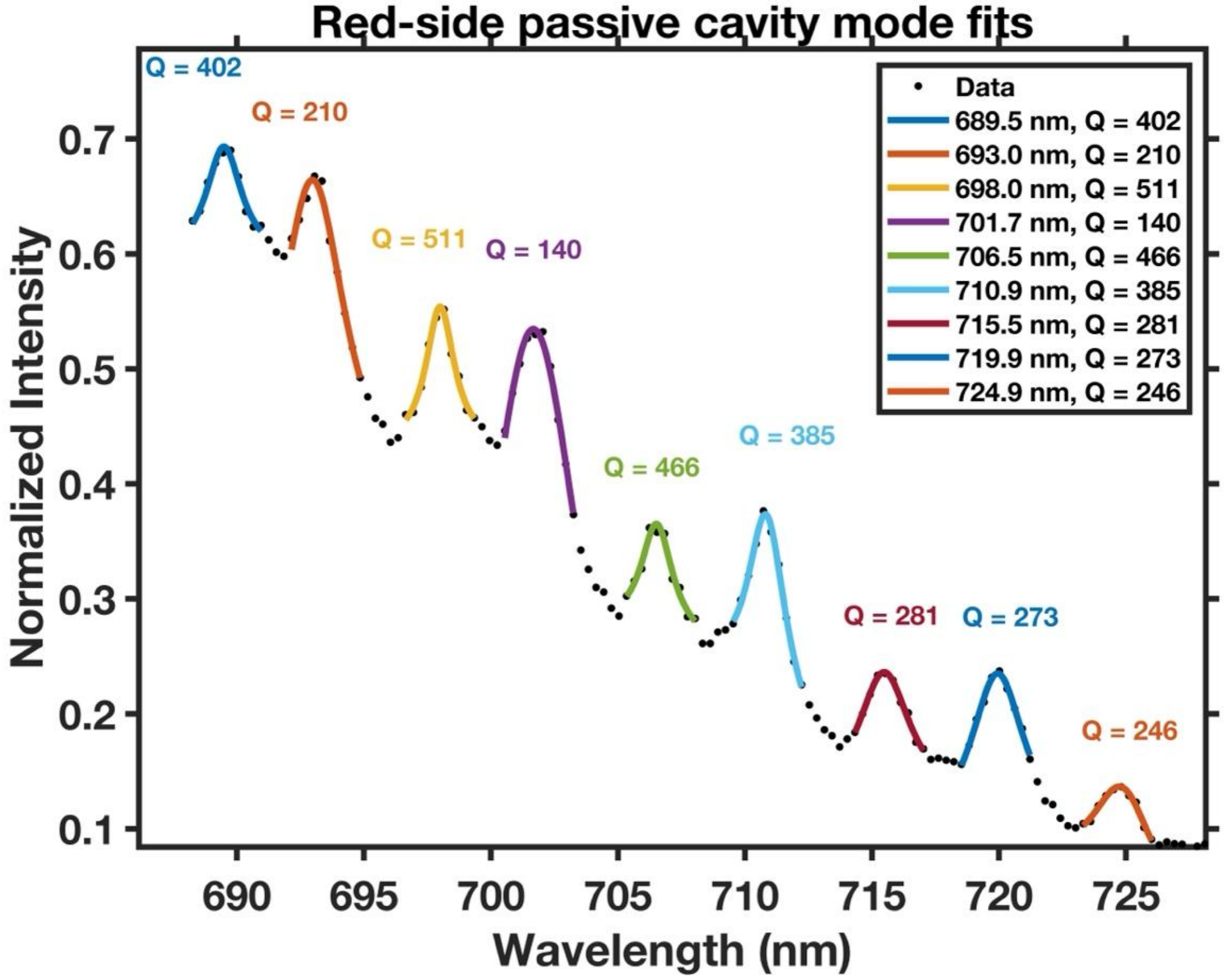

**Figure S9:** Passive quality factor analysis from near-infrared-excited photoluminescence. Red-side resonant features in a peripheral photoluminescence spectrum collected under continuous-wave 976 nm excitation were fit using Lorentzian functions with local linear backgrounds. The extracted mode center wavelengths and passive quality factors are indicated on the plot. These passive Q values estimate the cavity response under NIR-excited PL conditions and are distinct from active linewidth-derived Q values obtained above the lasing threshold.

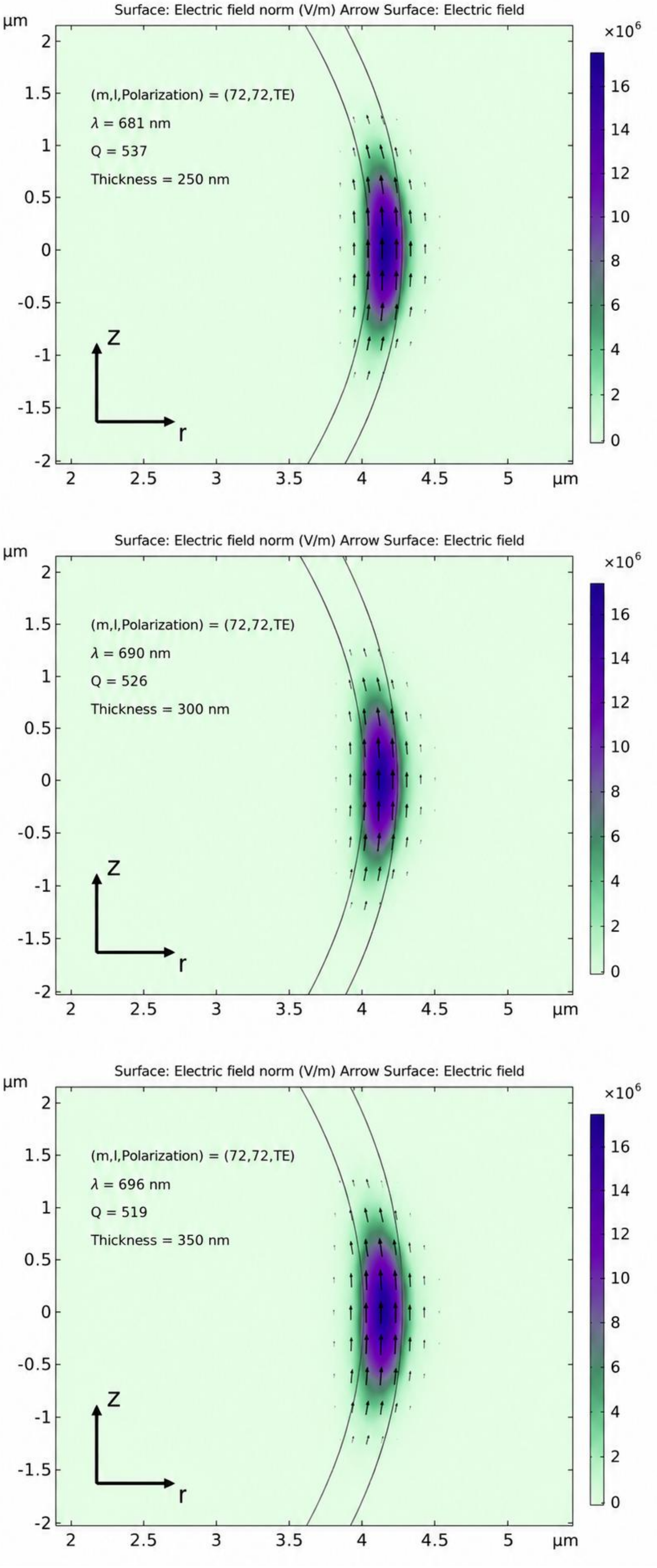
Surface: Electric field norm (V/m) Arrow Surface: Electric field
(m,l,Polarization) = (72,72,TE)
λ = 681 nm
Q = 537
Thickness = 250 nm
Surface: Electric field norm (V/m) Arrow Surface: Electric field
(m,l,Polarization) = (72,72,TE)
λ = 690 nm
Q = 526
Thickness = 300 nm
Surface: Electric field norm (V/m) Arrow Surface: Electric field
(m,l,Polarization) = (72,72,TE)
λ = 696 nm
Q = 519
Thickness = 350 nm
z
r
µm
×10^6

**Figure S10**: Numerical eigenmode calculations of shell-thickness effects in an ideal continuous colloidosome resonator**.** Electric-field magnitude and electric-field vectors are shown for the same TE-like $(m, l) = (72,72)$mode at shell thicknesses of (A) 250 nm, (B) 300 nm, and (C) 350 nm. The corresponding resonance wavelengths are 681, 690, and 696 nm, with calculated quality factors of 537, 526, and 519, respectively. The mode remains localized within and near the shell across the investigated thickness range. The calculations represent ideal smooth shells and do not include experimental losses associated with porosity, roughness, or QD-packing disorder; consequently, they do not establish an optimal or minimum shell thickness or directly predict the experimental threshold.

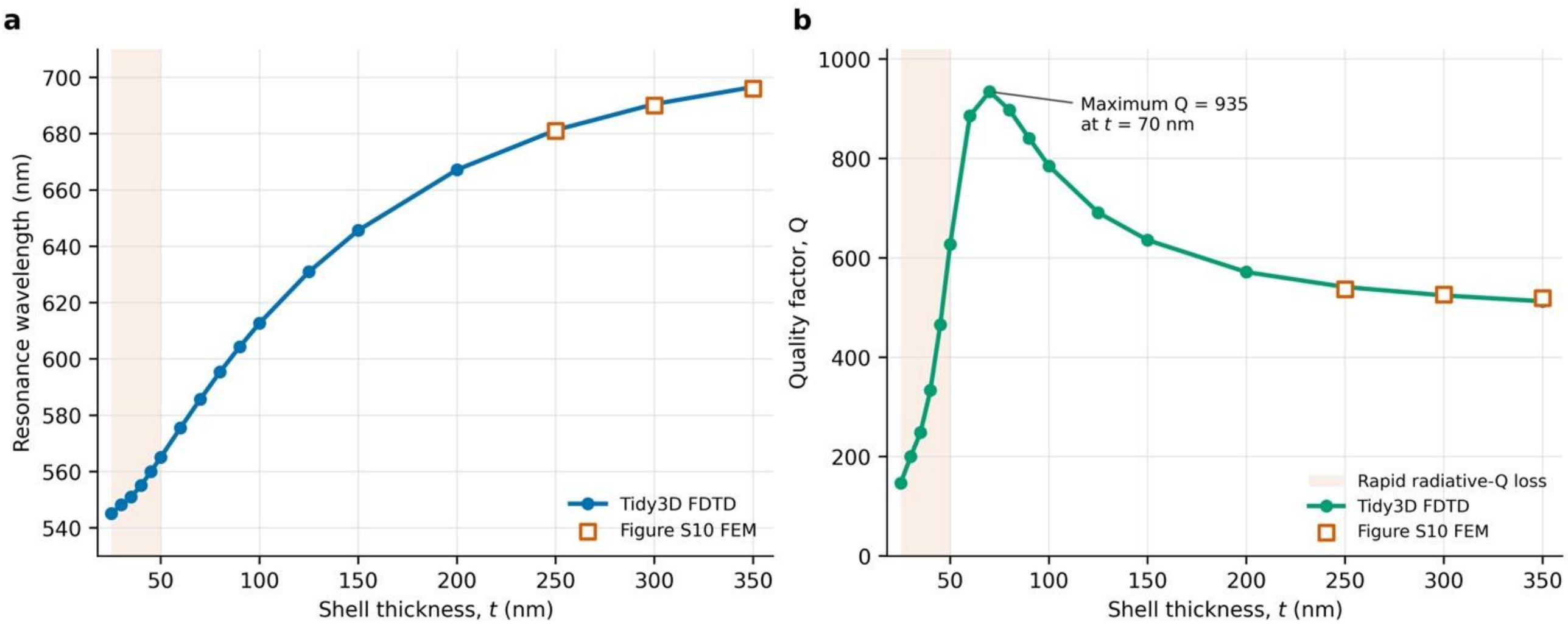

**Figure S11:** FDTD analysis of shell-thickness-dependent WGM behavior in an ideal continuous colloidosome. The outer particle radius and material parameters were held fixed while the shell thickness, t, was varied from 25 to 350 nm. (A) Resonance wavelength of the tracked m = 72 WGM as a function of shell thickness. The resonance red-shifts from 545.1 nm at t = 25 nm to approximately 695 nm at t = 350 nm. (B) Corresponding radiative quality factor, Q. The radiative Q increases rapidly through the thin-shell regime, reaches a maximum of Q = 935 at t = 70 nm, and then gradually decreases to approximately 500 across the 250–350 nm range. Orange squares show the corresponding FEM eigenmode results from Figure S10 at 250, 300, and 350 nm. The agreement between the FDTD and FEM calculations provides cross-method validation of the simulated thickness dependence.

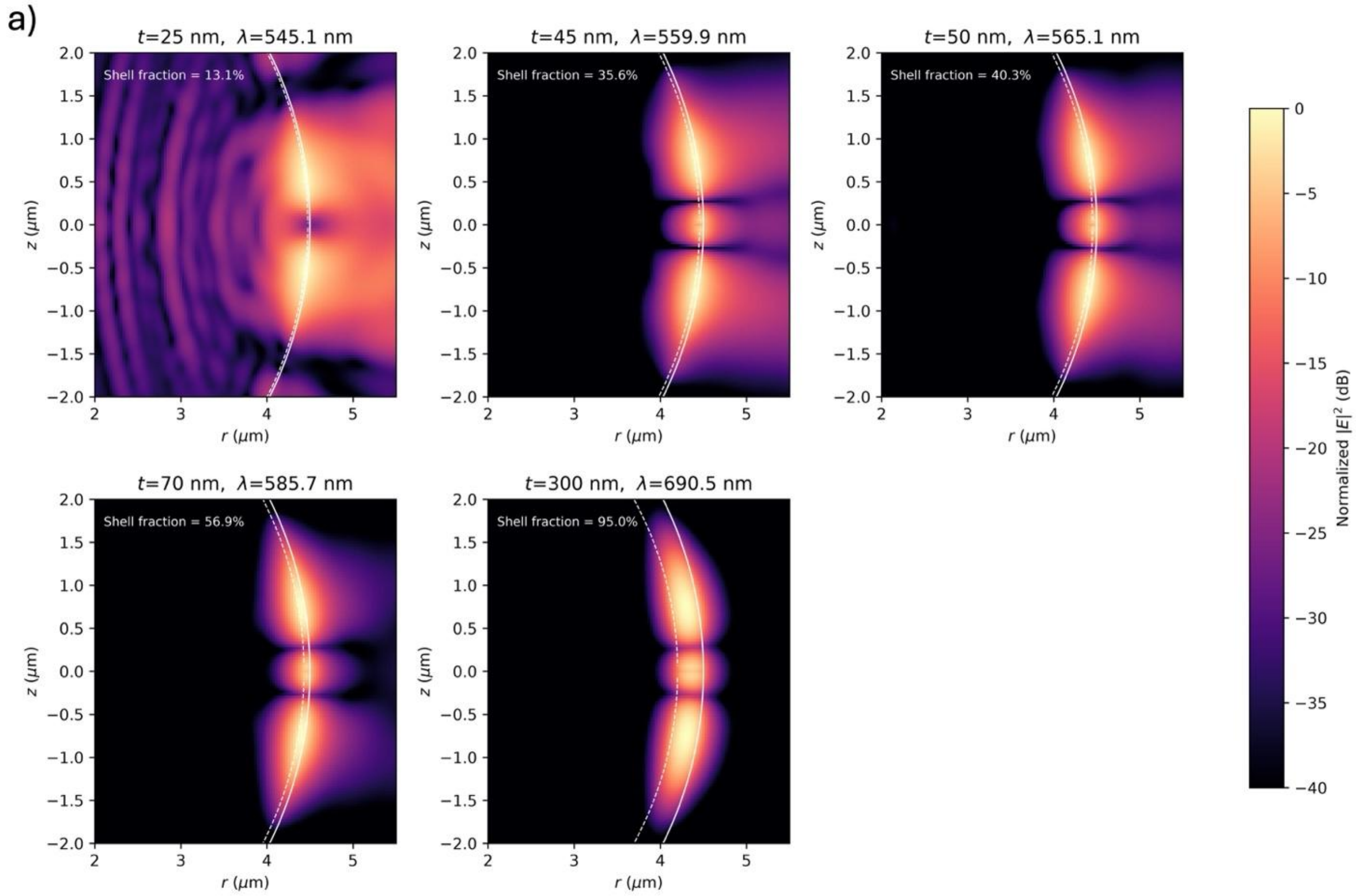

**Figure S12:** Representative FDTD field profiles showing the evolution of WGM shell localization with shell thickness. Meridional cross-sections of the normalized electric-field intensity, $|E|^2$, are shown on a decibel scale for the tracked m = 72 resonance at t = 25, 45, 50, 70, and 300 nm. The dashed curves indicate the inner and outer boundaries of the QD shell. At t = 25 nm, the mode is weakly localized to the shell, with a calculated shell-localization fraction of 13.1%. The shell-localization fraction increases to 35.6% at 45 nm, 40.3% at 50 nm, 56.9% at 70 nm, and 95.0% at 300 nm. The corresponding resonance wavelengths are 545.1, 559.9, 565.1, 585.7, and 690.5 nm, respectively. The field profiles demonstrate the continuous evolution from a weakly shell-localized thin-shell resonance to a strongly shell-confined WGM at the representative 300 nm thickness.

## Supporting references